\documentclass{spie}  
\usepackage[]{graphicx}
\usepackage{subcaption}

\title{Gemini planet imager observational calibrations X: non-redundant masking on GPI}

\author{Alexandra Z. Greenbaum\supit{a}, Anthony Cheetham\supit{b}, Anand Sivaramakrishnan\supit{c}, Peter Tuthill\supit{c}, Barnaby Norris\supit{c}, Laurent Pueyo\supit{b}, Naru Sadakuni\supit{d}, Fredrik Rantakyr\"o\supit{d}, Pascale Hibon\supit{d}, Stephen Goodsell\supit{d}, Markus Hartung\supit{d}, Andrew Serio\supit{d}, Andrew Cardwell\supit{d}, Lisa Poyneer\supit{e}, Bruce Macintosh\supit{e,f}, Dmitry Savransky\supit{g}, Marshall D. Perrin\supit{b}, Schuyler Wolff\supit{a}, Patrick Ingraham\supit{f}, Sandrine Thomas\supit{h} with the GPI team.
\skiplinehalf
\supit{a}The Johns Hopkins University, 3600 N. Charles St., Baltimore, MD, USA 
\skiplinehalf
\supit{b}University of Sydney, School of Physics, University of Sydney, NSW 2006, Australia 
\skiplinehalf
\supit{c}Space Telescope Science Institute, 3700 San Martin Drive, Baltimore MD 21218 USA  
\skiplinehalf
\supit{d}Gemini Observatory, Casilla 603, La Serena, Chile
\skiplinehalf
\supit{e}Lawrence Livermore National Lab, 7000 East Ave., Livermore, CA 94551, USA 
\skiplinehalf
\supit{f}Kavli Institute for Particle Astrophysics and Cosmology, Stanford University, Stanford, CA 94305, USA 
\skiplinehalf
\supit{g}Sibley School of Mechanical and Aerospace Engineering, Cornell University, Ithaca NY, USA 14853.
\skiplinehalf
\supit{h}UARC/NASA Ames Research Center, MS 244-10, Moffett Field, CA 94035, USA }

\authorinfo{Further author information: (Send correspondence to A.Z.G.)\\A.Z.G.: E-mail: agreenba@pha.jhu.edu, Telephone: 1 410 338 2429}

\begin{document} 
  \maketitle 

\begin{abstract}
The Gemini Planet Imager (GPI) Extreme Adaptive Optics Coronograph contains an
interferometric mode: a 10-hole non-redundant mask (NRM) in its pupil wheel.
GPI operates at \textit{Y, J, H}, and \textit{K} bands, using an integral field
unit spectrograph (IFS) to obtain spectral data at every image pixel.  NRM on
GPI is capable of imaging with a half resolution element inner working angle at
moderate contrast, probing the region behind the coronagraphic spot.  The fine
features of the NRM PSF can provide a reliable check on the plate scale, while
also acting as an attenuator for spectral standard calibrators that would
otherwise saturate the full pupil.  NRM commissioning data provides details
about wavefront error in the optics as well as operations of adaptive optics
control without pointing control from the calibration system. We compare lab and
on-sky results to evaluate systematic instrument properties and examine the
stability data in consecutive exposures.  We discuss early on-sky
performance, comparing images from integration and tests with the
first on-sky images, and demonstrate resolving a known binary. We discuss the
status of NRM and implications for future science with this mode.   
\end{abstract}


\section{INTRODUCTION}
The Gemini Planet imager is an extreme AO coronagraph \cite{macintosh2014first}
with a 10-hole non-redundant mask (NRM) in its pupil
\cite{2010SPIE.7735E.266S}. GPI's extreme AO wavefront control
\cite{poyneer2006, poyneerthis} uses a Shack-Hartmann wavefront sensor upstream
of the coronagraph and a calibration system (CAL), a low-order wavefront sensor
measuring slow tip/tilt and other modes relative to the coronagraphic focal
plane mask\cite{2003ESASP.539..513L, 2008SPIE.7015E.180W} (the
high-order sensing capability of CAL is not active). Imaging modes using the
unocculted science mirror (direct and NRM) do not allow light to pass to the
CAL for these low order corrections. Commissioning the instruments has seen a
series of on and off sky tests of the various imaging modes and the adaptive
optics controls with and without tip/tilt corrections from the CAL system. 

NRM provides interferometric resolution, to angular scales of about $\lambda /
2B$, $B$ being the longest hole-pair distance, at high dynamic range
($10^2-10^3$ binary contrast routinely possible from the ground). This makes NRM a
well suited technique for imaging hot planet forming regions inside young
circumstellar disks with GPI. In particular, NRM is a powerful probe of gaps in
transition disks that may be harboring the building blocks of planets
\cite{2012ApJ...745....5K, 2013ApJ...762L..12C, 2012ApJ...753L..38B}.  
GPI provides spectral information by employing an integral field unit
spectrograph (IFS). Imaging in spectral mode may enable the detection of
emission features associated with strong circum-planetary accretion in these
regions. NRM can also operate with the imaging polarimeter
\cite{2010SPIE.7736E.192P} and may provide an eye into polarized disks at small
inner working angles.

NRM images are also a diagnostic tool for the operation of the instrument and
wavefront correction controls. For IFS images, NRM fringes are a sensitive
measure of the pixel scale relative to the NRM mask and can provide an
independent check on the wavelength calibration if the plate scale (i.e., the
magnification from the telescope) is known \cite{2013SPIE.8864E..1VG}.  The
sensitivity of the NRM PSF, due to improved angular resolution, could also be a
good tool to measure atmospheric dispersion and dispersion correction,
especially in brighter sources.

\label{sec:intro}  


\section{The GPI NRM }

\begin{figure}[htbp!] \centering \begin{frame}{ \includegraphics[height=3cm
\vspace*{5mm}]{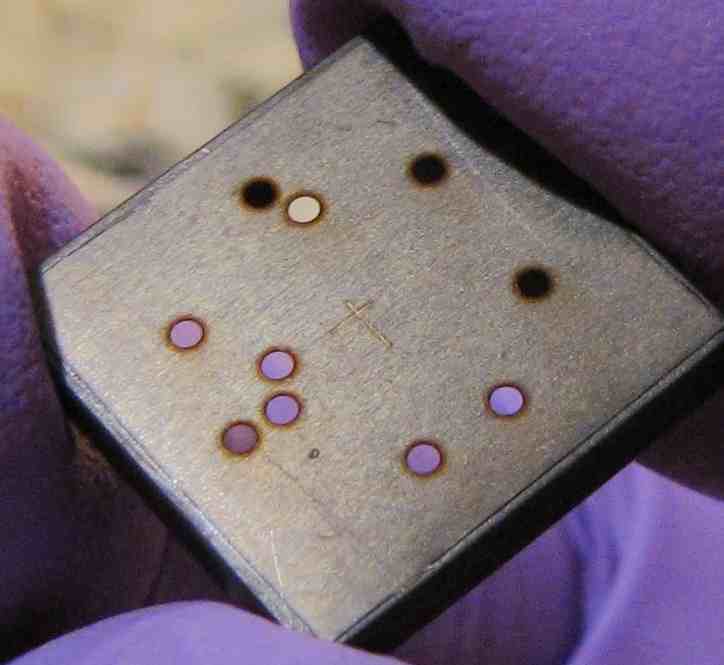}} \end{frame}
\includegraphics[height=5.5cm]{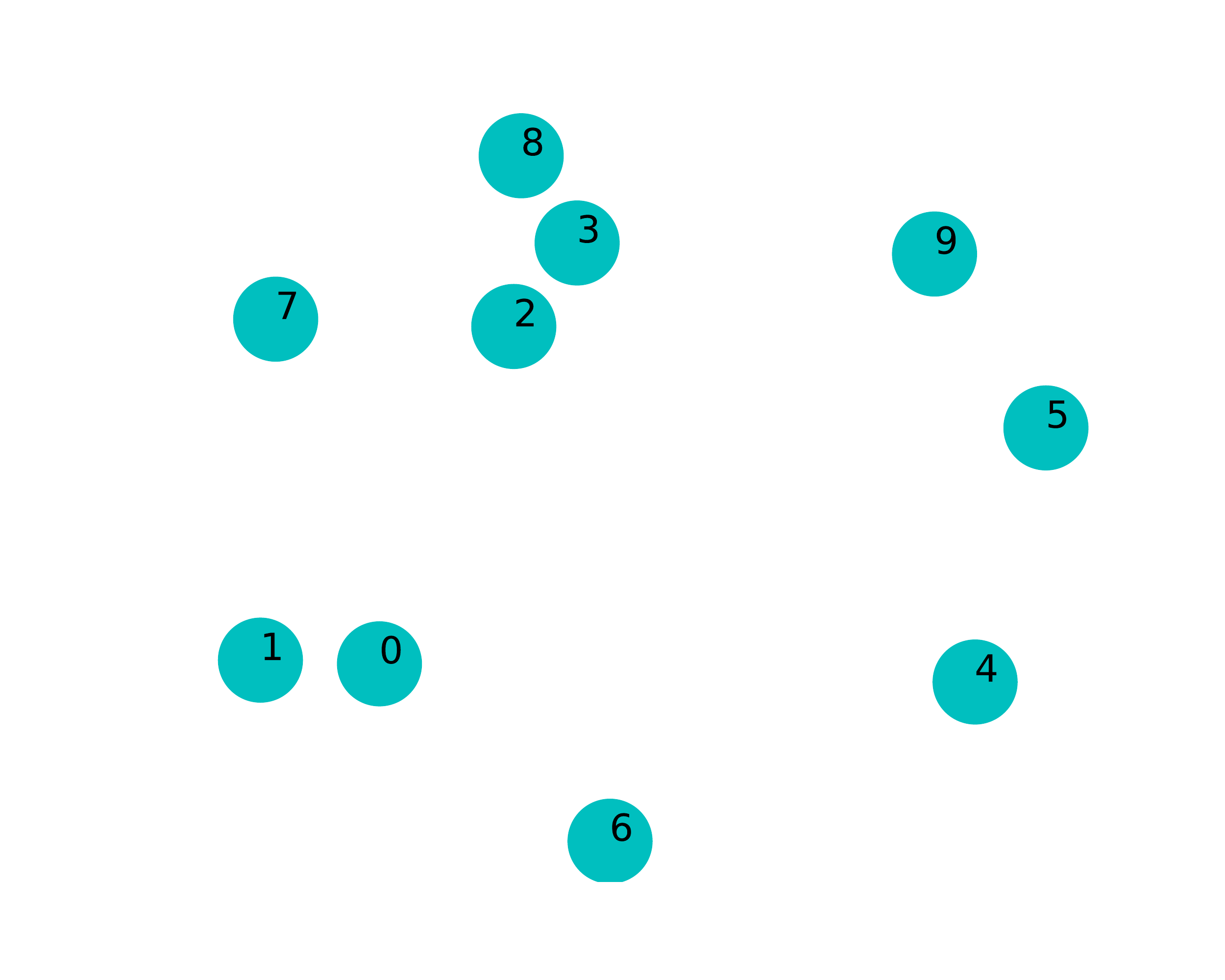}
\includegraphics[height=5.5cm]{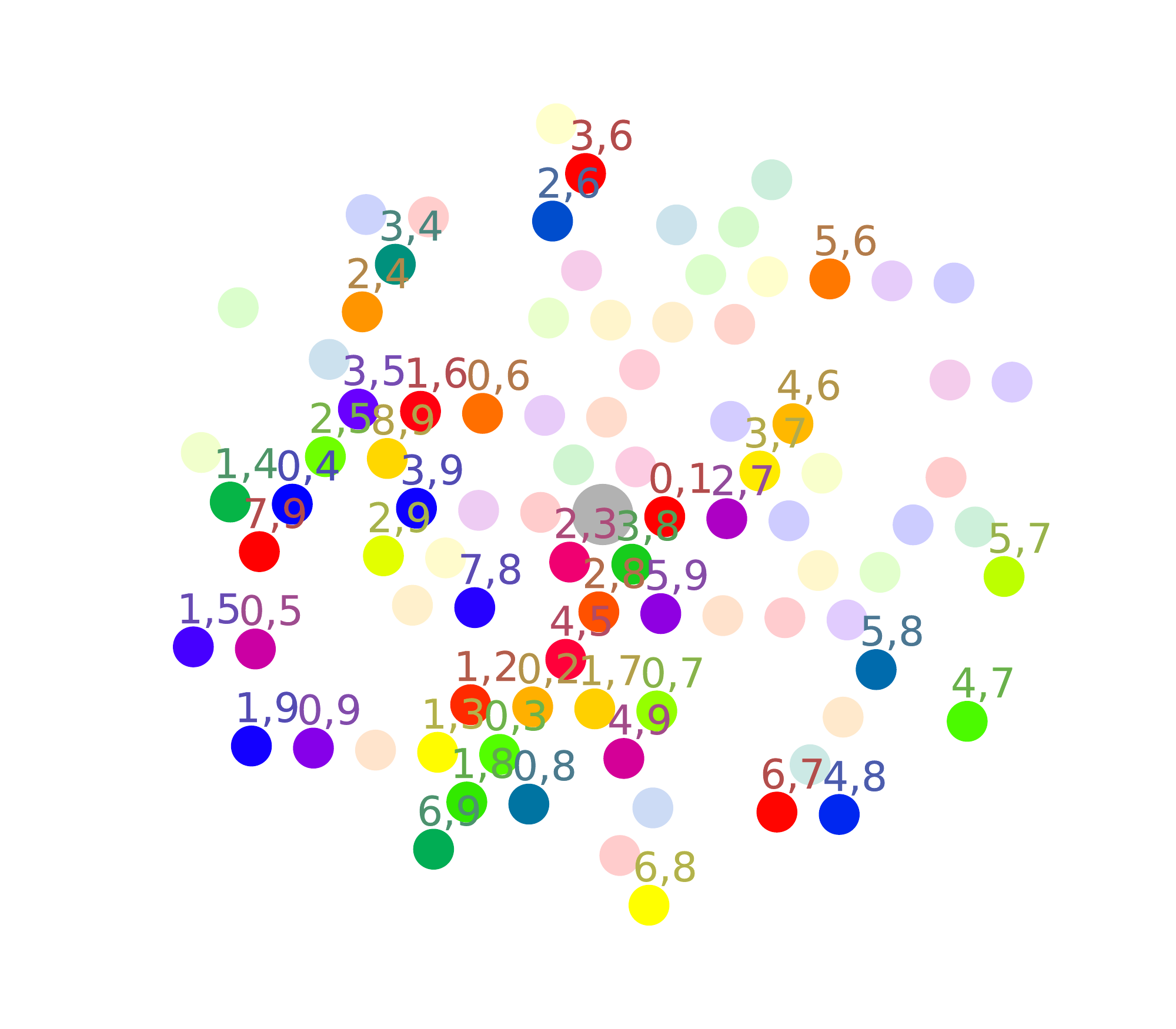}
\caption{\label{fig:labeled} \textbf{Labeled Holes and Splodges}: \textbf{Left:} GPI's 10-hole
NRM\cite{Lenox}. \textbf{Center:} We establish a numbering convention, which helps
to keep track of hole and baseline dependent systematics and aids diagnosis of
problems, shown in the center. \textbf{Right:} we plot the associated hole-pair
baselines.} \end{figure}

GPI's NRM has 10 holes that form 45 non-redundant baseline vectors. The mask is
designed to provide good coverage in spatial frequencies without strong directional
preference.  The middle of Figure \ref{fig:labeled} shows our hole labeling convention that
we will refer to throughout this paper. The right of Figure \ref{fig:labeled} shows the
corresponding hole-pair baselines plotted in the Fourier plane (spatial
frequencies).


Measuring fringe phases and amplitudes from point sources is a powerful
diagnostic for the instrument and provides information regarding the wavefront.
The fringe stability over subsequent exposures determines performance for
imaging science with NRM. Careful analysis of the fringes, both through Fourier
transform diagnostics and an analytic fringe fitting can reveal the angular
scale of various instrument instabilities.

In this paper we focus on the NRM observations from GPI commissioning in the
context of diagnosing instrument systematics and instabilities. In Section
\ref{sec:analysistools} we describe how to analyze the data in Sections
\ref{sec:alignment} and \ref{sec:observations} we describe alignment procedures
and observations. In Section \ref{sec:diagnostics} we explore the details of
diagnostics NRM provides to GPI in general. Section \ref{sec:performance}
provides a demonstration of contrast and presents analysis of early contrast
detection performance that will be reported more comprehensively in future
communications.

   \begin{figure}
   \centering
   \begin{tabular}{l c}
   \includegraphics[width=0.8\textwidth]{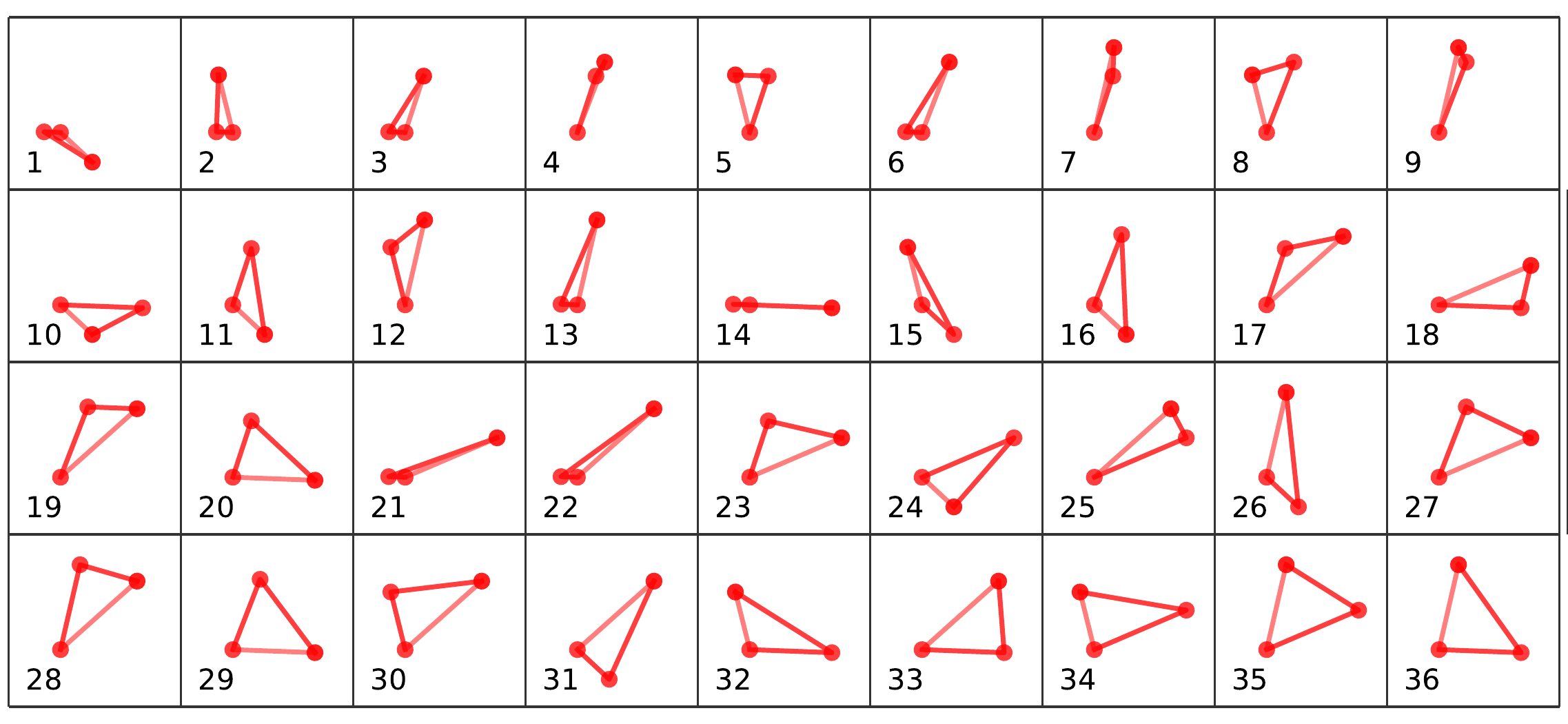}   
   \end{tabular}
   \caption 
   { \label{fig:triangles} \textbf{Independent Closure Triangles:}
A set of 36 unique closure triangles in order of increasing perimeter. The addition of
fringe phases around these closing triangles is insensitive to piston phase error in 
the pupil.}
   \end{figure} 

\section{Analysis Techniques } 
\label{sec:analysistools}
Fringe amplitudes and phases are extracted from NRM images through Fourier
principles. These fringe observables reveal information about the wavefront.
The fringe amplitudes can be measured by fringe contrast and fringe phases by
fringe center shift in the image. For $N$ holes there are $N(N-1)/2$ baselines.
Phases measure antisymmetric structure, appropriate for detecting multiple
point source objects and faint companions. Amplitudes can measure
centro-symmetric structure. Measuring fringe observables may be done with a
good model of the fringes that form the image, or by Fourier transforming the
data, both to measure signal in spatial frequencies that correspond to mask
baselines. 

The closure phase is the sum of fringe phases around a closed triangle.
Imaging a point source with relatively stable and unaberrated optics, this sum
should be zero for all $N\choose 3$ closure triangle combination
\cite{1958MNRAS.118..276J, 1986Natur.320..595B, 1987Natur.328..694H}.  Figure
\ref{fig:triangles} displays an example of a subset of 36 independent closure
triangles, in general $(N-1)(N-2)/2$, for GPI's mask, where all baselines are
represented. 
Closure phases have the remarkable property of immunity to any piston phase
errors located in a pupil 
Non-zero closure phase can be the result of structure in the image or
uncorrected high order wavefront error due to atmosphere or imperfect optics.
Precision in the closure phase measurement sets the limit on achievable
contrast with NRM. 

\subsection{Analytic Fringe Fitting} 

We fit fringe phases with an analytic model of the NRM PSF that describes the
PSF as a a sum of fringes corresponding to each hole-pair baseline
\cite{2011A&A...532A..72L,Greenbaumsubmitted}.  The NRM PSF can be represented
mathematically by an envelope, $P$ (the Airy pattern for circular holes as in
GPI NRM), modulated by sinusoidal fringes from each baseline.  Constant pistons
can be expressed as a constant phase shift in the transform of the of the pupil
mask. For pupil units $\b{x}$ and image units $\b{k}$:

\begin{eqnarray}  \label{eq:PSF}
   a(\b{k})a^{*}(\b{k}) = P(\b{k})  \sum_{i=1}^N \sum_{j=1}^N  e^{- i \b{k}
\cdot( \b{x}_{i}-\b{x}_{j}) +  i(\phi_{i} - \phi_{j})} 
\end{eqnarray}
In general, relative pistons between holes contribute
constant fringe phases $\Delta\phi_{i,j}$'s at hole centers $\b{x}_i$'s.
\begin{eqnarray} \label{eq:model} \nonumber psf =P(\b{k})\{10+ \cos{(
\b{k}\cdot(\b{x}_{1}-\b{x}_{2}))}\cos{(\Delta\phi_{1,2})} \\ \nonumber %
-\sin{( \b{k}\cdot(\b{x}_{1}-\b{x}_{2}))}\sin{(\Delta\phi_{1,2})}  \\ \nonumber
+ \cos{( \b{k}\cdot(\b{x}_{1}-\b{x}_{3}))}\cos{(\Delta\phi_{1,3})} \\ \nonumber
-\sin{( \b{k}\cdot(\b{x}_{1}-\b{x}_{3}))}\sin{(\Delta\phi_{1,3})} \\ +...\}
\label{eq:model}
\end{eqnarray}
The model can be tuned to reflect plate scale, sub-pixel centering, and other
similar parameters. These can be fit for in the Fourier plane. For the GPI data
cubes, we consider each slice to be monochromatic in the model.

\subsection{Fourier approach and diagnostic tools}

We also provide an independent analysis of the data through the Fourier
approach \cite{2000PASP..112..555T, 2006ApJ...650L.131L, 2013MNRAS.433.1718I}
with the aperture masking pipeline at University of Sydney.  Additionally,
Fourier transforming the data provides a quick and intuitive look directly at
the phase and amplitude behavior with time or wavelength and can help diagnose
baseline-dependent errors. Tracking down the hole pairs that contribute to
phase peculiarities can help distinguish between hole-dependent errors and
baseline-dependent errors, and also reveal morphological signatures in the
phase. 

\section{Alignment in Pupil } \label{sec:alignment}
\begin{figure}[!htbp] 
\centering 
\includegraphics[height=4cm]{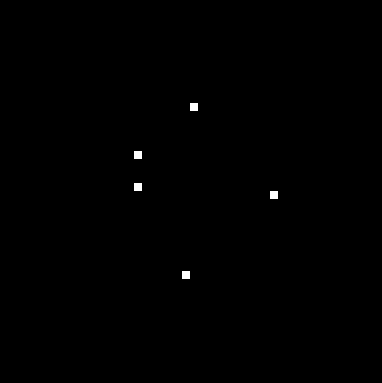}
\includegraphics[height=4cm]{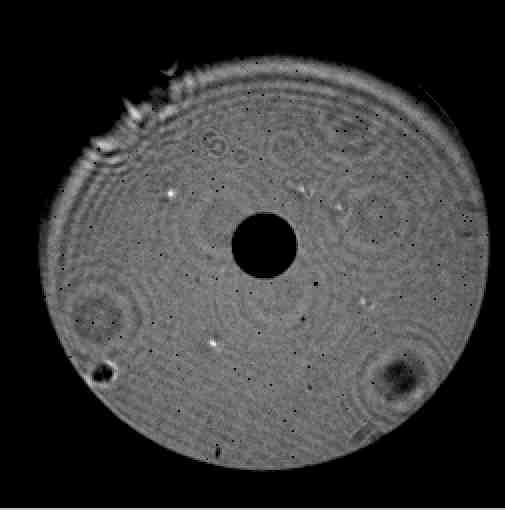}
\caption {\textbf{Left}: asymmetric 5-poke pattern on the $48 \times 48$ GPI
MEMS actuator array. \textbf{Right:} The resulting pupil image. The pupil image
is vignetted at the top, a result of miaslignment  of the pupil-sensing camera,
not the science path. The pupil viewing camera also shows other features like
dust on the camera and bad actuators.  \label{fig:asymmpoke} } 
\end{figure}

The first thing to check when using NRM with extreme AO is mask alignment in
the pupil.  We determined the mask position in the pupil with a few custom poke
patterns on the deformable mirror, illuminating the pupil with the internal
source, imaging with the pupil viewing camera. One asymmetric set of pokes
helped determine the rotation between the the MEMS plane (Figure
\ref{fig:asymmpoke}) and the pupil to help map the NRM holes to actuator
locations. The rotation between the poke pattern and the as-designed NRM
orientation (with respect to the MEMS DM) is roughly 243.6 degrees
counter-clockwise Another set confirmed a new aligned position of the mask.

\begin{figure}[!htbp]
\centering
\includegraphics[height=4cm]{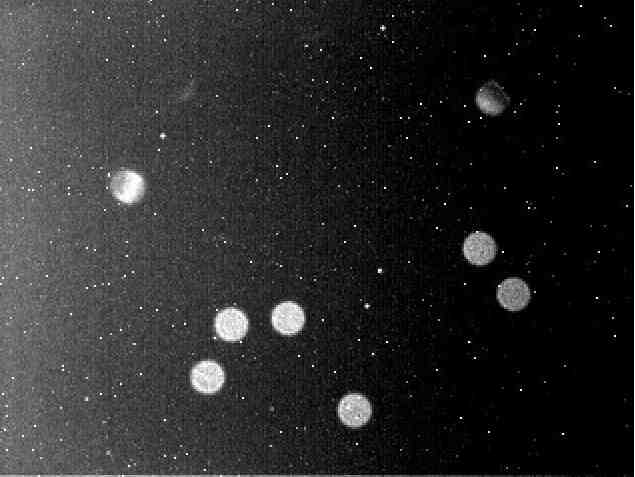}
\includegraphics[height=4cm]{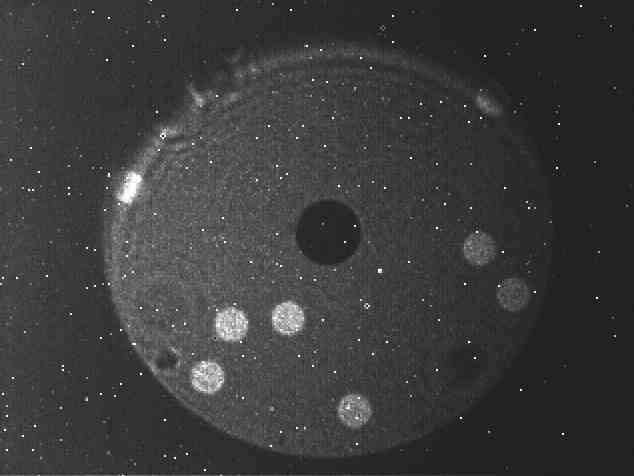}
\caption {\textbf{Left:} The nominal alignment of the NRM seen with the pupil
viewing camera. Holes 4 and 5 look completely cut off, hole 6 partially.
\textbf{Right:} Pupil viewer images of the NRM and overlaid with the clear
pupil show vignetting by the pupil viewing system. \label{fig:nrm_unaligned} }
\end{figure}

\begin{figure} 
\centering
\includegraphics[height=4cm]{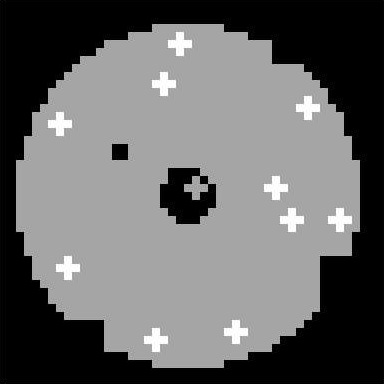}\\
\includegraphics[height=4cm]{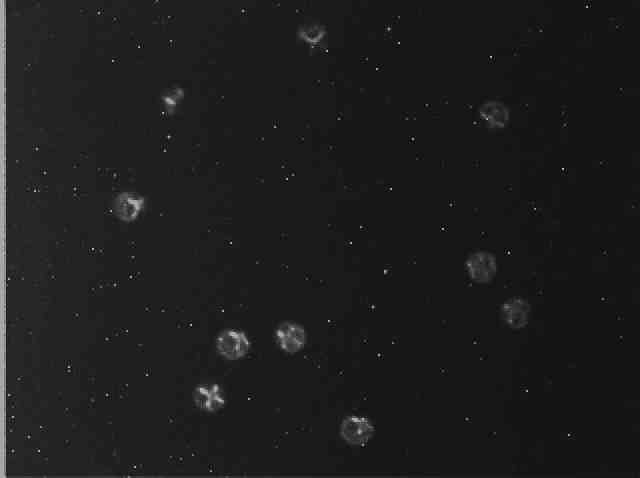}
\includegraphics[height=4cm]{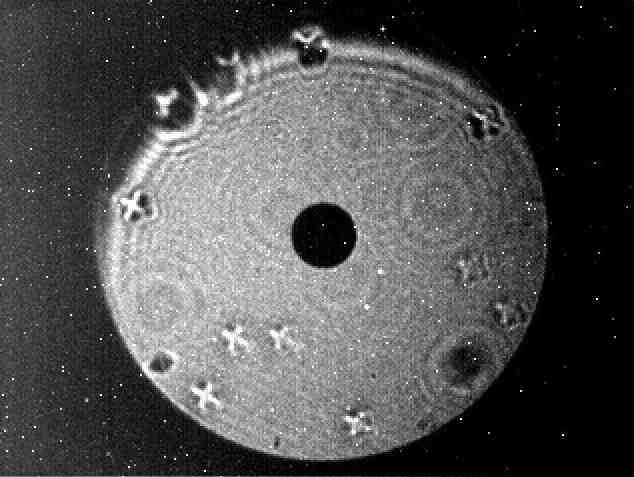} 
\caption {\label{fig:nrmmap2badact} \textbf{Top:} The expected NRM hole centers
(white crosses) overlaid on the bad actuator map of active subapertures (gray).
\textbf{Bottom left:} The poke pattern seen through the mask moved to its new
position.  \textbf{Bottom right:} The poke pattern seen through the clear
pupil. The NRM hole centers miss known bad actuator locations.} 
\end{figure}

Figure \ref{fig:nrm_unaligned} shows the NRM in its nominal position from the
pupil viewer (a) and overlaid on a clear pupil image (b). A few holes appear to
be clipped, but since the pupil viewer optics vignet the on-axis beam it is
difficult to determine the edge. The secondary obstruction provided a good
handle on the pupil center and which holes were cut off at the edge. We shifted
the mask from the nominal position to lie entirely in the pupil. Figure
\ref{fig:nrmmap2badact} shows a poke pattern of NRM hole centers mapped to the
deformable mirror actuators, and as seen through the mask in the pupil. A set
of NRM-specific pokes (Figure \ref{fig:nrmmap2badact}) confirmed the new hole
position mapping to the MEMS plane and confirmed our alignment. The actuators
covered by each hole are obvious and can be used to track back peculiarities in
the AO control. We confirm that the holes miss spiders and known dead actuator
locations, as designed. The dead actuators are indicated in masked regions at
the top of Figure \ref{fig:nrmmap2badact}.


\section{Observations } 
\label{sec:observations}
NRM observations on GPI were taken over the course of several commissioning
runs. These have provided a good sample of both short exposure and longer
exposure point source observations as operations changes throughout
commissioning. The observations discussed in this manuscript are summarized in
Table \ref{table:obs}.

\begin{table}[!htbp]
\centering
\caption{\label{table:obs} Summary of commissioning observations discussed in this report}
\begin{tabular}{|l || l l | l | l |}
\hline
Month 	& Target description 	& Details & Purpose & Notes\\ \hline\hline
Dec 	& - HR 2690 (binary)\cite{hartkopf2012} & 6 frames in $H$ for 60s & Performance &The first complete NRM dataset\\
2014	& - HR 2716 (calib) & 6 frames in $H$ for 60s &verification& saw long exposures that produced\\ 
	& - HR 2839 (calib) & 6 frames in $H$ for 60s && inconsistent phases between frames. \\ 
	& 	& && Seeing was 0.6" or better. \\ \hline
Mar 	& &  &Engineering& We aimed for shorter exposures to \\
2014	& - HD 63852 (pt src)& 20 frames $Y,H$ for 1.5s && better understand instabilities. \\ 
	& 	& 		&&Seeing was 0.6" or better. \\ \hline
May 	& Point sources  &&& We Tested new AO control  \\
2014	& - HD 63852	& 20 frames in $H$ for 1.5s &Engineering \&& software with short and long\\
	& - HD 142695 	& 8 frames in $J$ for 54s &Performance& exposures on known point sources\\
	& - HD 142384	& 8 frames in $J$ for 37s & verification& used for NRM calibration \cite{2012ApJ...753L..38B} and took\\
	& 		& &&  a sequence of short exposures with \\
	& Internal source & 60 frames in $H$ at 1.5s&& the internal source. Median seeing \\
	&  	&  	&& was 0.8".\\
\hline
\end{tabular}
\end{table}

\section{Instrument diagnostics } \label{sec:diagnostics}
We have identified the most obvious systematics in the data. These are a
combination of static phase and amplitude trends and temporal instabilities
that ultimately limit contrast and resolution with GPI NRM. Understanding the
behavior of the instrument in NRM can help measure the instrument wavefront in
static conditions (e.g. lab or internal source measurements) as well as
diagnose unseen behaviors in the adaptive optics control that operates without
CAL tip/tilt correction. CAL must receive light through the coronagraph hole to
send to a low-order shack hartman sensor, with a 4 pixel camera (quad cell) to
measure tip/tilt in the calibration system \cite{2003ESASP.539..513L,
2008SPIE.7015E.180W}. All non-coronagraphic science exposures operate without
this low order correction.

\subsection{Static phases}
\begin{figure}[!htbp]
\centering
\includegraphics[height=5cm]{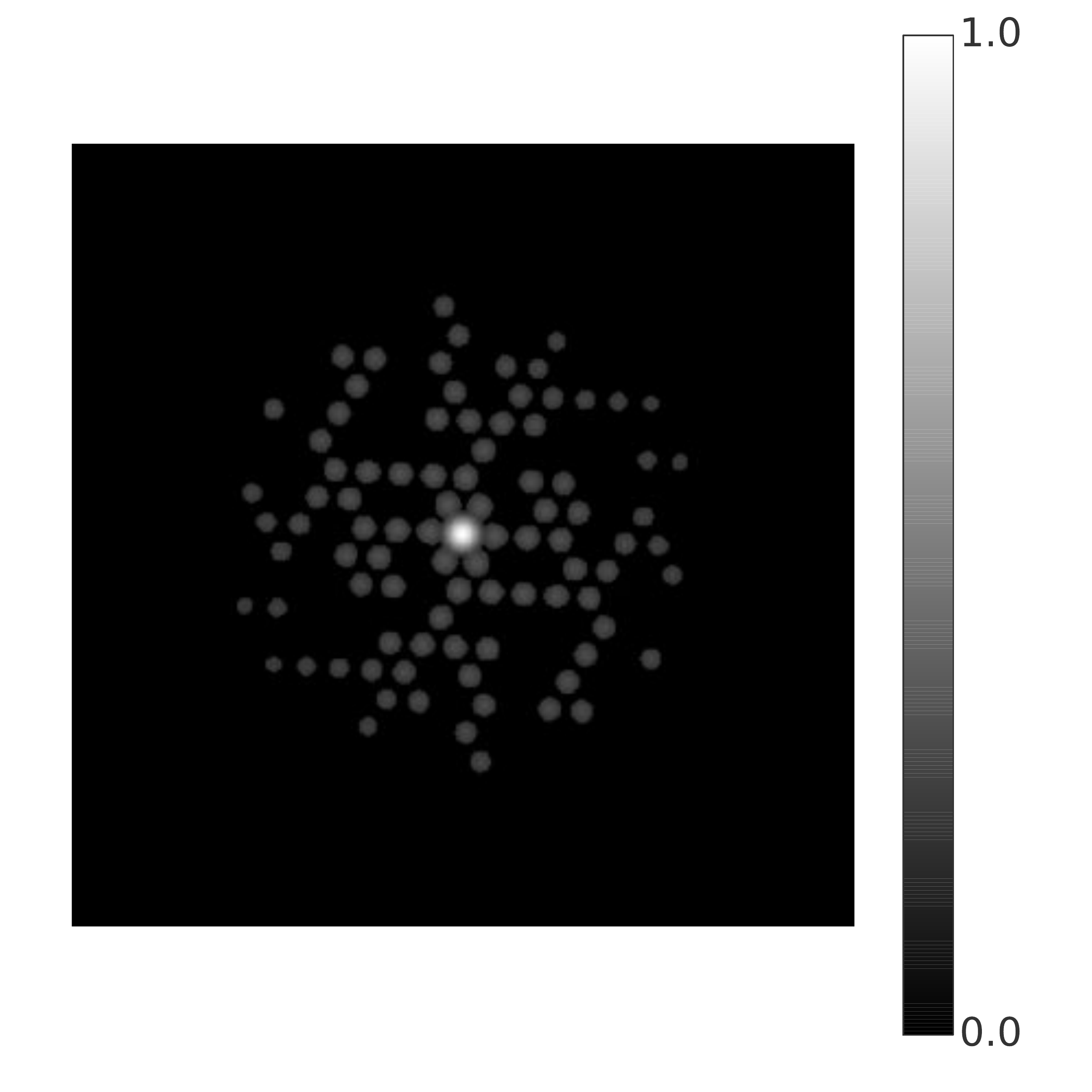}
\includegraphics[height=5cm]{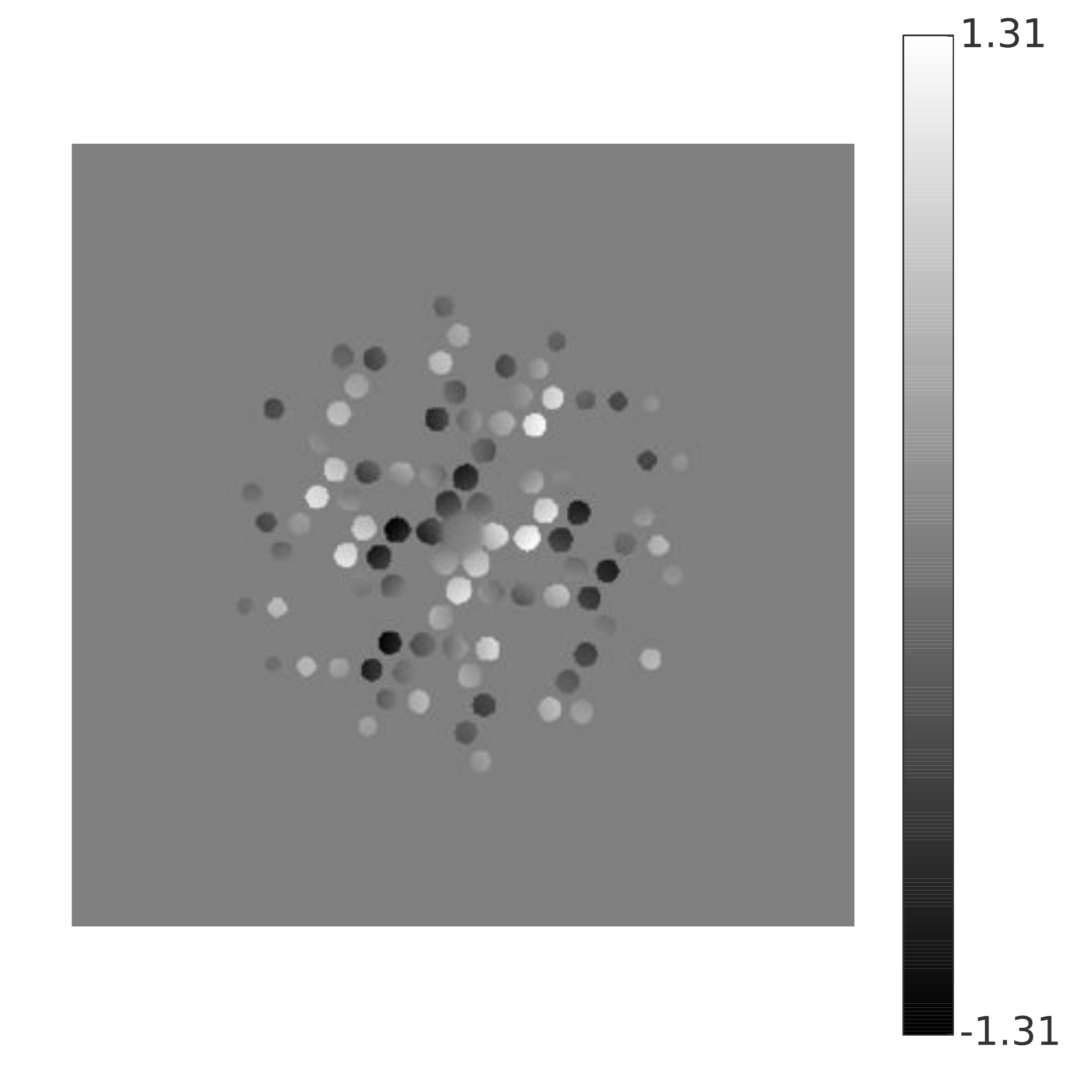} \\
\includegraphics[height=5cm]{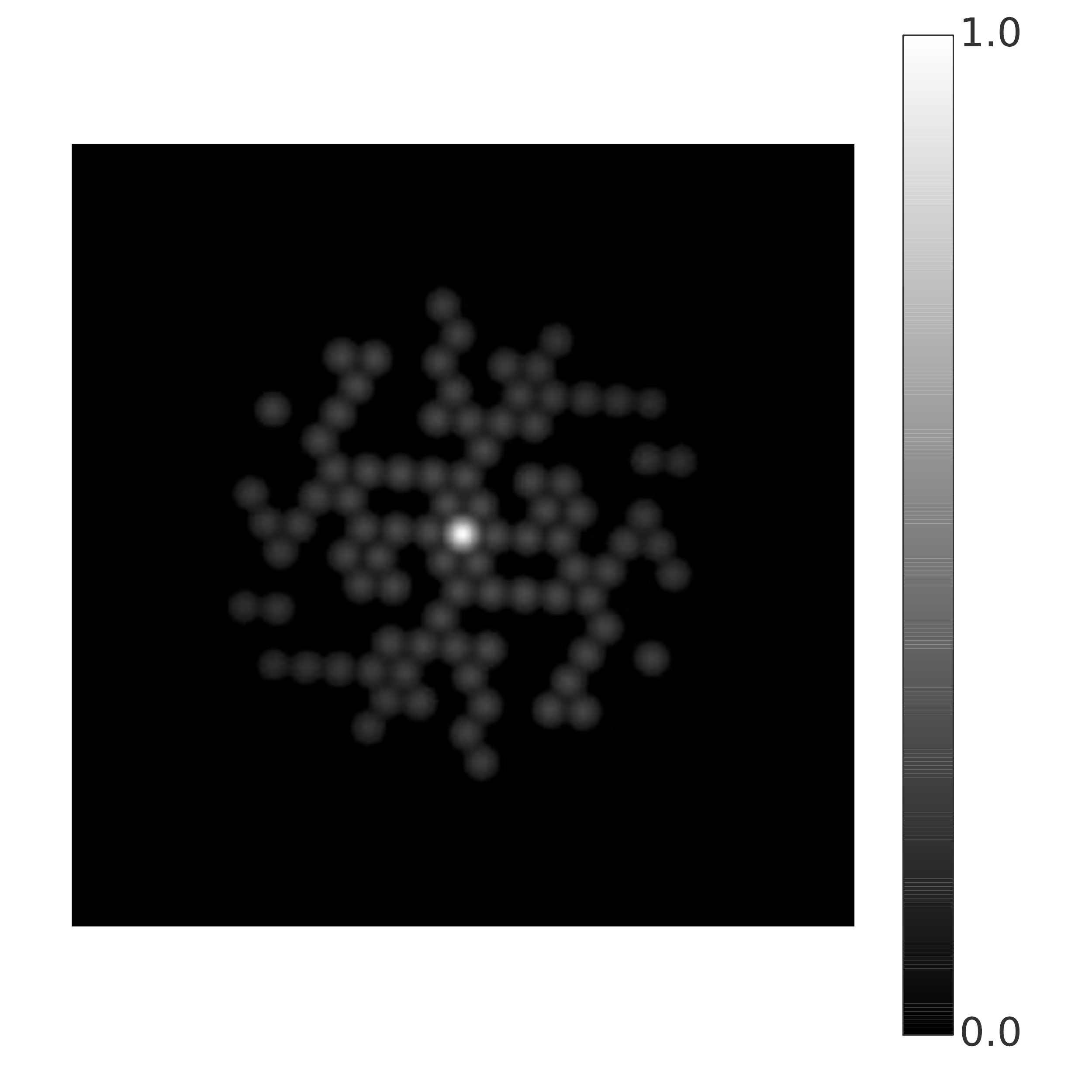}
\includegraphics[height=5cm]{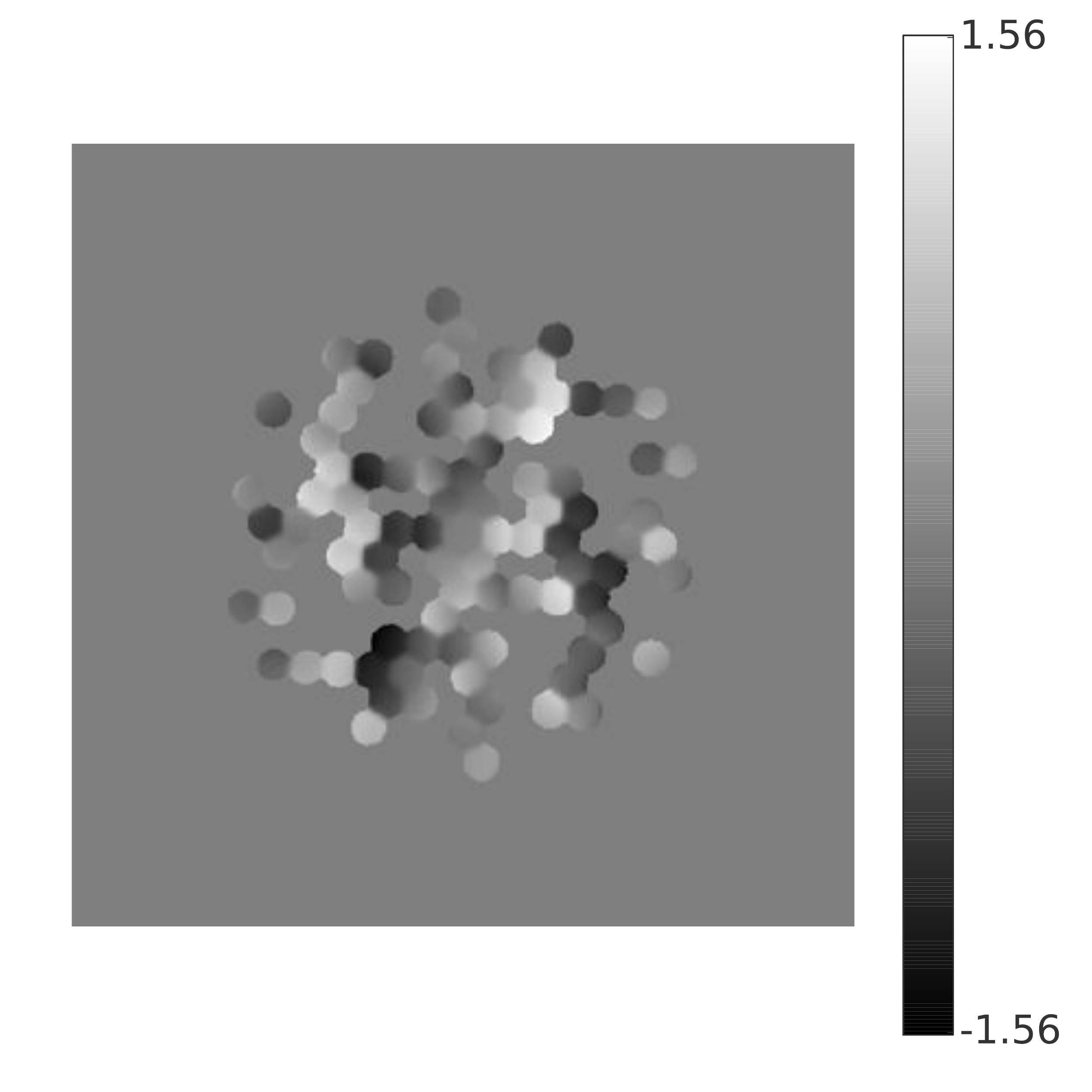}
\caption{\label{fig:labphases} \textbf{Top:} Fringe amplitudes (Left) and
phases (Right) measured from July 2013 integration and tests from an average of
several exposures.  \textbf{Bottom:} Amplitudes and phases from May 2014
internal source single exposure on telescope.  Both sets of amplitudes drop
with longer baselines. The phases appear  morphologically similar. }
\end{figure}

\begin{figure}[!htbp]
\centering

\includegraphics[height=6cm]{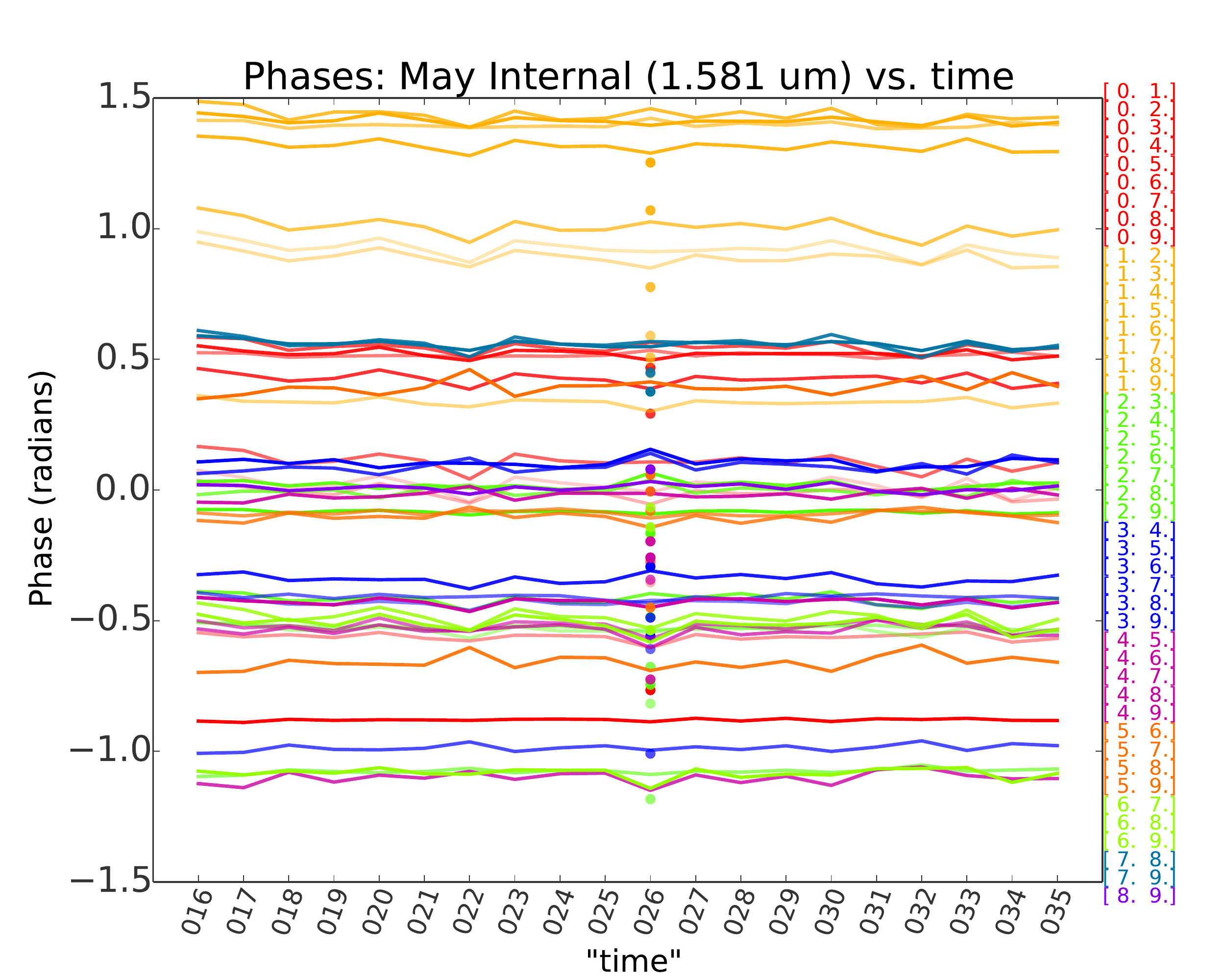}
\caption{\label{fig:mayphases} May 2014 internal source phases plotted over
time.  The single data-points correspond to average phase from July 2013
integration and tests.  They are in general lower in July when the instrument
was not attached to the telescope.}
\end{figure}

We observe static piston phases in the NRM fringes during the most stable
exposures in the lab (before commissioning) and with the internal source while
on the telescope. Static pistons do not affect NRM contrast because they
calibrate out in the closure phase calculation. However, they provide
information about the instrument wavefront after AO correction. 
Exposures from the stable internal source (measured by analytic model fitting)
match the general trend of phases measured during integration and tests in July
2013.  Figure \ref{fig:labphases} shows the morphology of the fringe phases
matches between July 2013 and May 2014. We plot the phases for each baseline in
Figure \ref{fig:mayphases} over 20 exposures in May (solid lines) compared to
those recovered from July intergration and tests (dots).

Static phases seen by the NRM are measuring the wavefront inherent in the
optical system, possibly caused by the non-common-path of the AO system and the
imaging detector. Such wavefront error can cause quasi-static speckles that can
limit the obtainable contrast of coronagraphy and conventional imaging, but
calibrate out in closure phase. Knowing the static phases could potentially
feed into the adaptive optics control and improve contrast in other imaging
modes. 


\subsection{Wavelength dependent instabilities}
December 2013 commissioning data was taken with the NRM misaligned in the
pupil, so that holes 4,5, and 6 were potentially clipped by the edge of the
pupil, which could contribute to decreased amplitudes in related baselines. Any
wavefront error remaining after AO correction will cause wavelength dependent
changes in both amplitude and phase. If there is Fresnel fringing, for example
from near field vignetting in the optical path, we may also expect wavelength
dependent effects.  We note that odd behavior at the edges of the band in our
measurements are a result of reduced filter throughput at band edges
\cite{mairethis}.

\begin{figure} [!htbp]
\centering
\includegraphics[height=6cm]{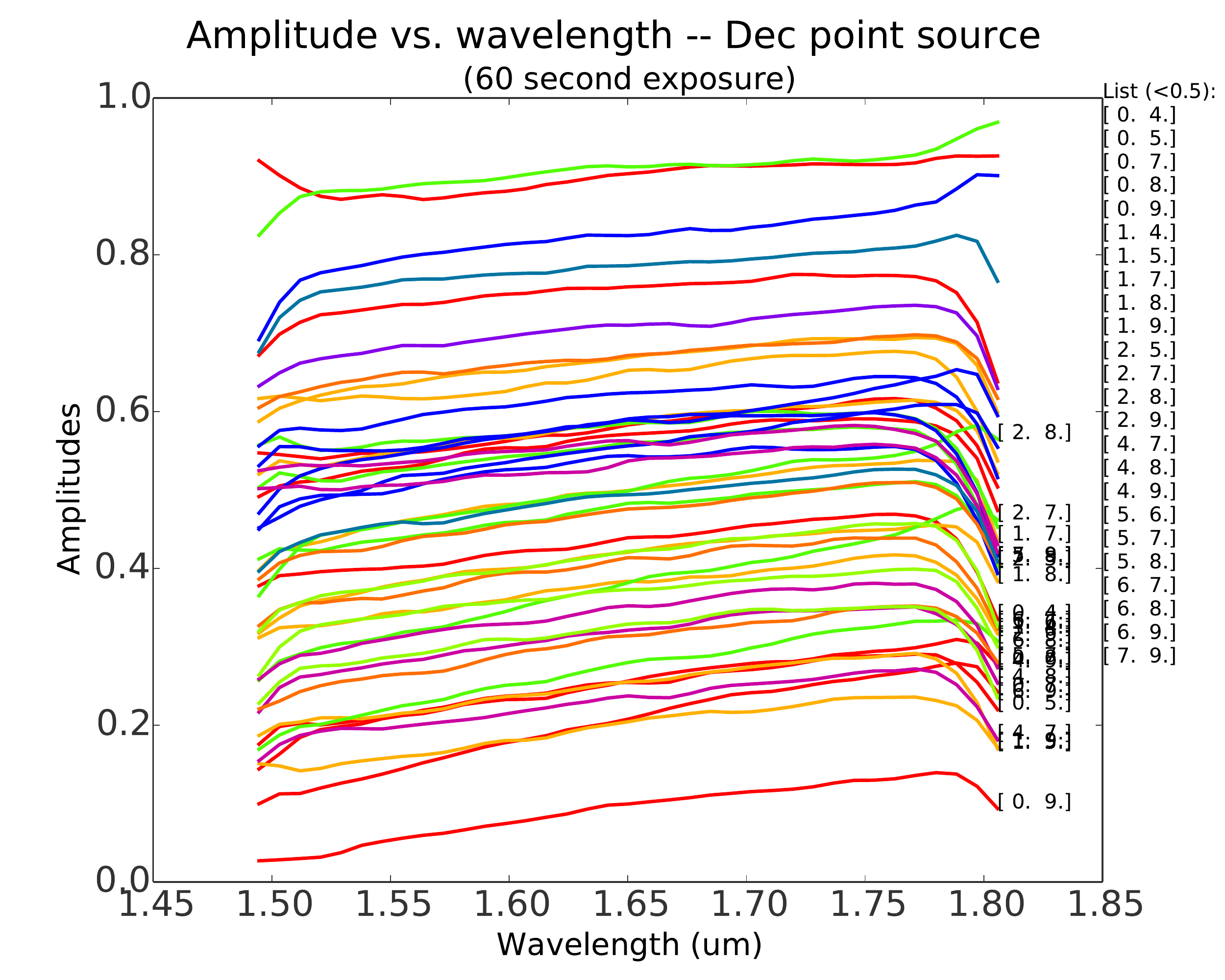}
\includegraphics[height=6cm]{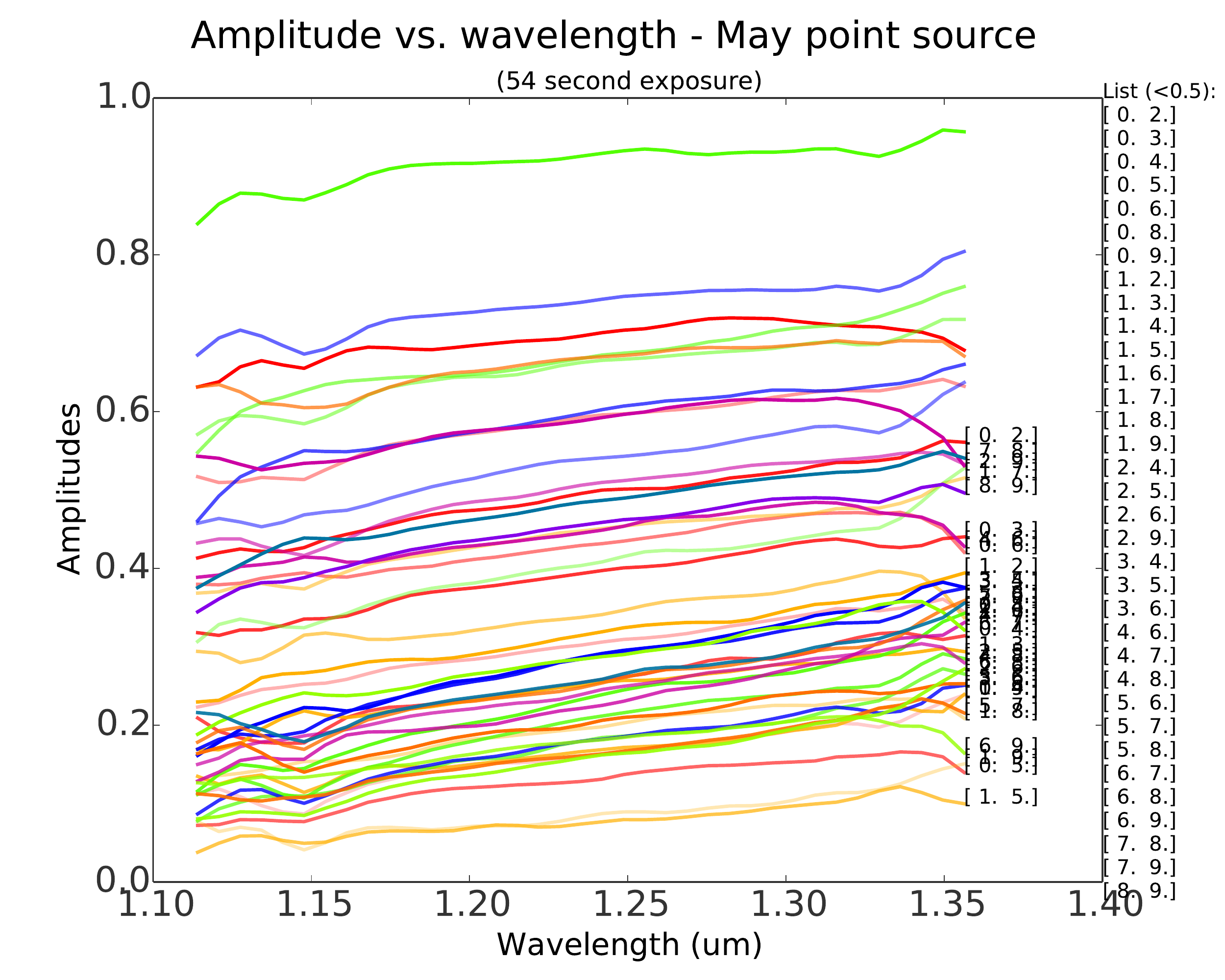}
\includegraphics[height=6cm]{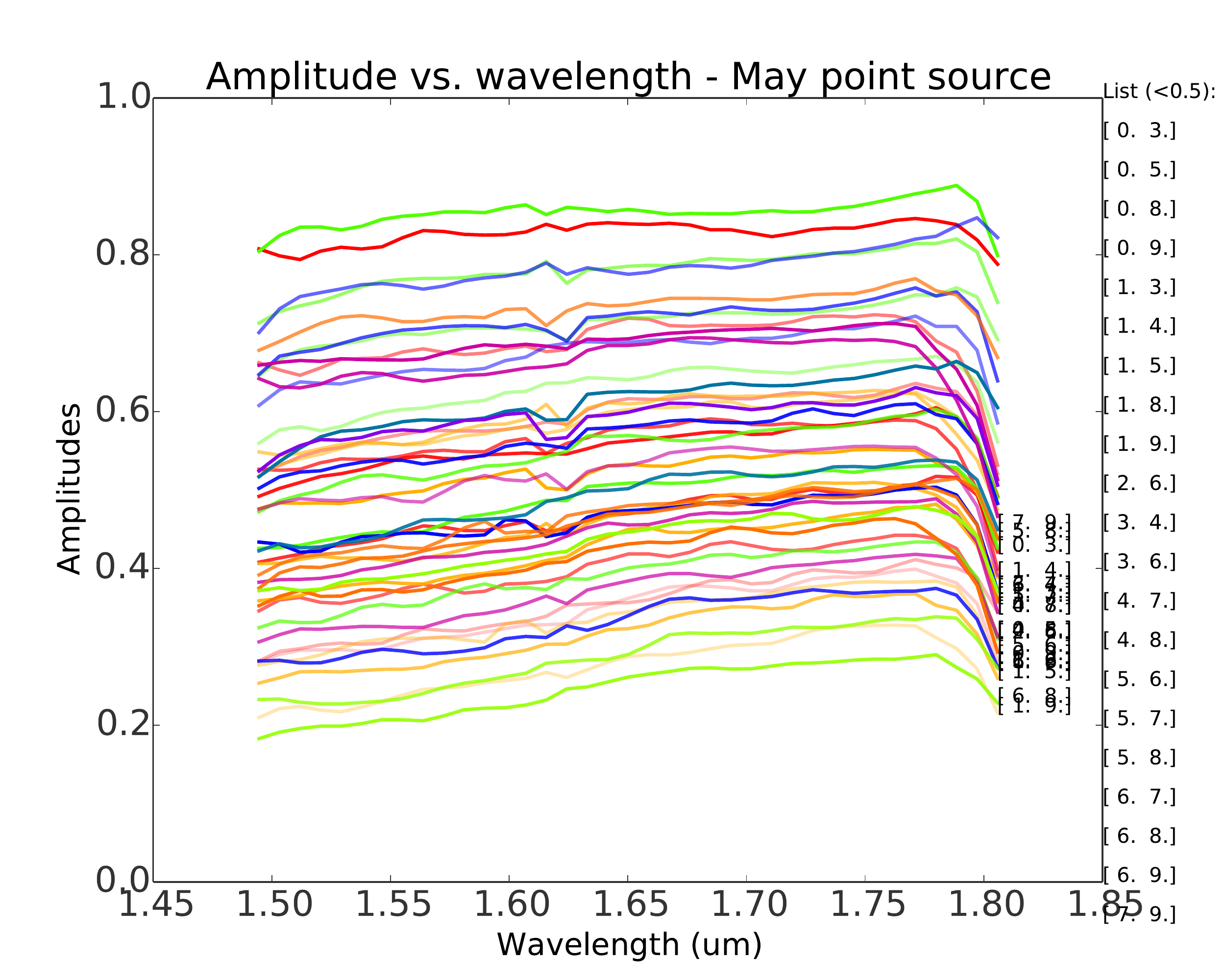}
\caption {\label{fig:ampvwl} Average fringe amplitude over an observing
sequence for each wavelength slice. Baselines [1,5] and [0,9] show up
consistenly between observations. Low amplitudes at longer baselines are 
at least partially due to jitter in the image.} 
\end{figure}

In all datasets we see the amplitudes rise with increasing wavelength (Figure
\ref{fig:ampvwl}), especially in lower average amplitude baselines, which tend
to be longer. This is true even for data taken with the internal source while
on the telescope and integration and test data from the lab at UCSC. The trends
are relatively flat at the higher visibility amplitude baselines, and become
more steeply sloped at longer baselines where wavefront error has the largest
effect. The effect is more pronounced in on-sky data. 

Comparing amplitudes from exposures during integration and testing to internal
source exposures on the telescope shows what additional instabilities exist on
the telescope from various vibrations. The rise in amplitude with longer
wavelength exist in the I\&T data, which also shows the lower amplitude
baselines that involve holes 9 and 5 seen in many of the on-sky datasets.  The
effect systematically limits fringe amplitude measurements at some of the
longer baselines and at shorter amplitudes. This may explain improved contrast
at longer wavelengths.  Currently, vibration and non-common path error are the
best candidates for the source of reduced visibility amplitude at longer
baselines.  Residual wavefront error should scale with wavelength, just as we
observe.

\begin{figure} [!htbp]
\centering \includegraphics[height=6.5cm]{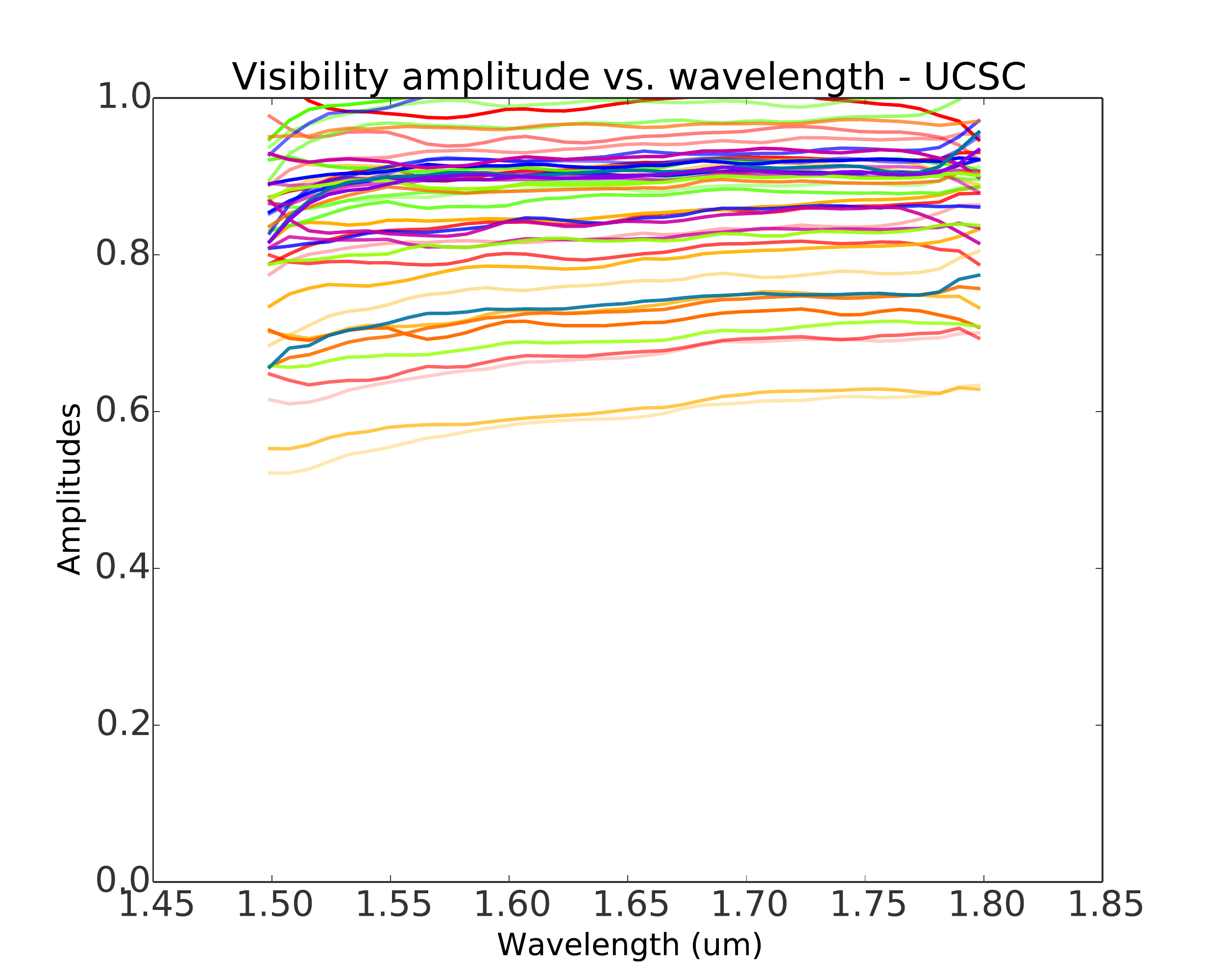}
\includegraphics[height=6.5cm]{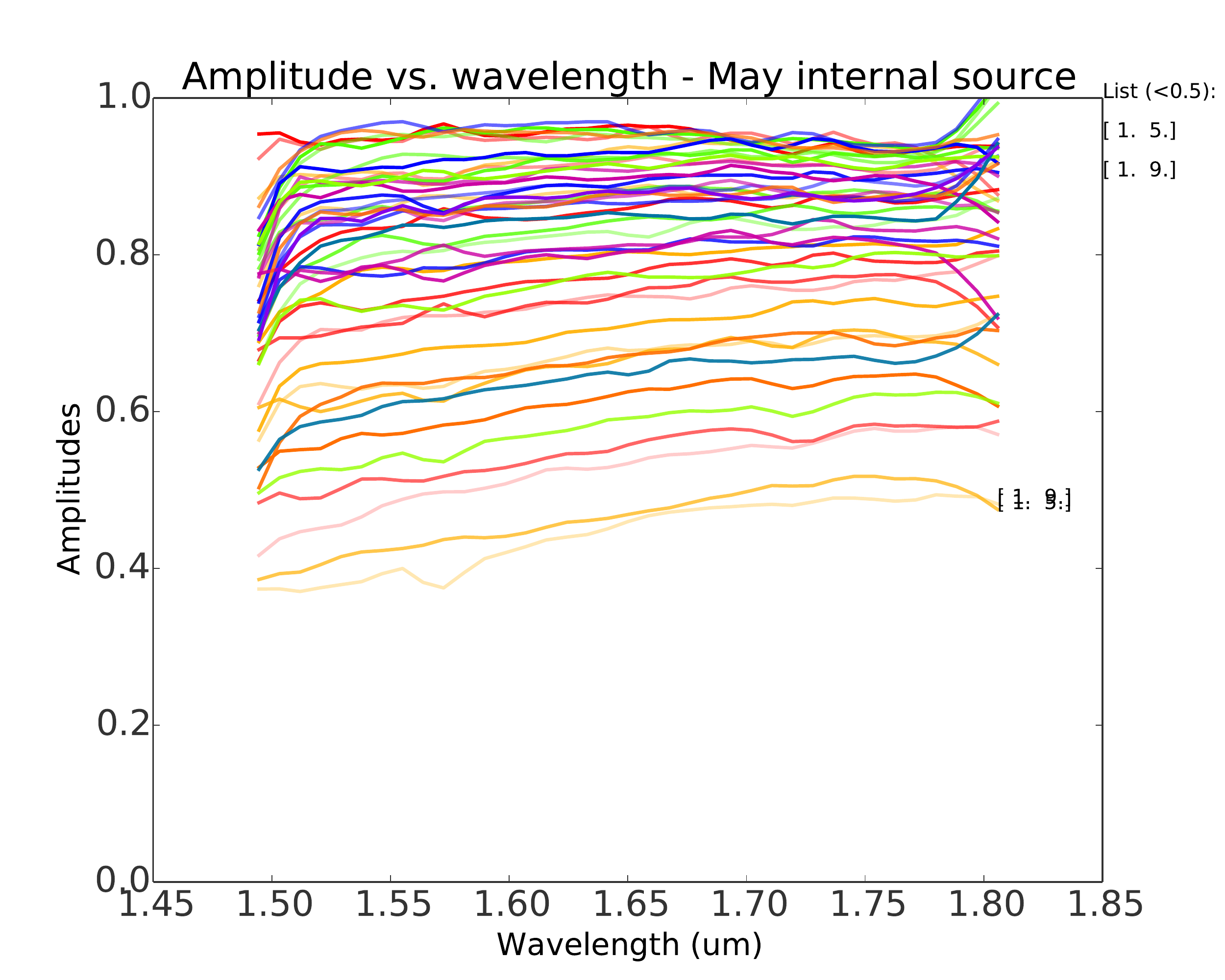}
\caption {\label{fig:ampvwl_lab} Visibility amplitude behavior with wavelength
for an mean exposure from integration and test data (left) and the mean over a
set of 20 exposures with the internal source in May. The two display a similar
trend.  Particular baselines show lower amplitudes and there is systematic
wavelength dependence, possibly due to instrument vibrations. The amplitudes
are worse while the instrument is on the telescope, which could be a result of
additional vibrations.} 
\end{figure}

\subsection{Pointing instabilities}
Data taken in December 2013 showed significant drops in amplitude, especially
in some longer baselines, indicating some kind of temporal instability.  In
March we took a set of short exposures ($1.5$s) on point source HD 63852 to see
how fringes behaved on shorter timescales. While displaying the sharpest
fringes on sky, these revealed jitter in the image, within about a pixel, close
to $14 \mathrm{mas}$.
\begin{figure} [!htbp]
\centering
\includegraphics[height=6.5cm]{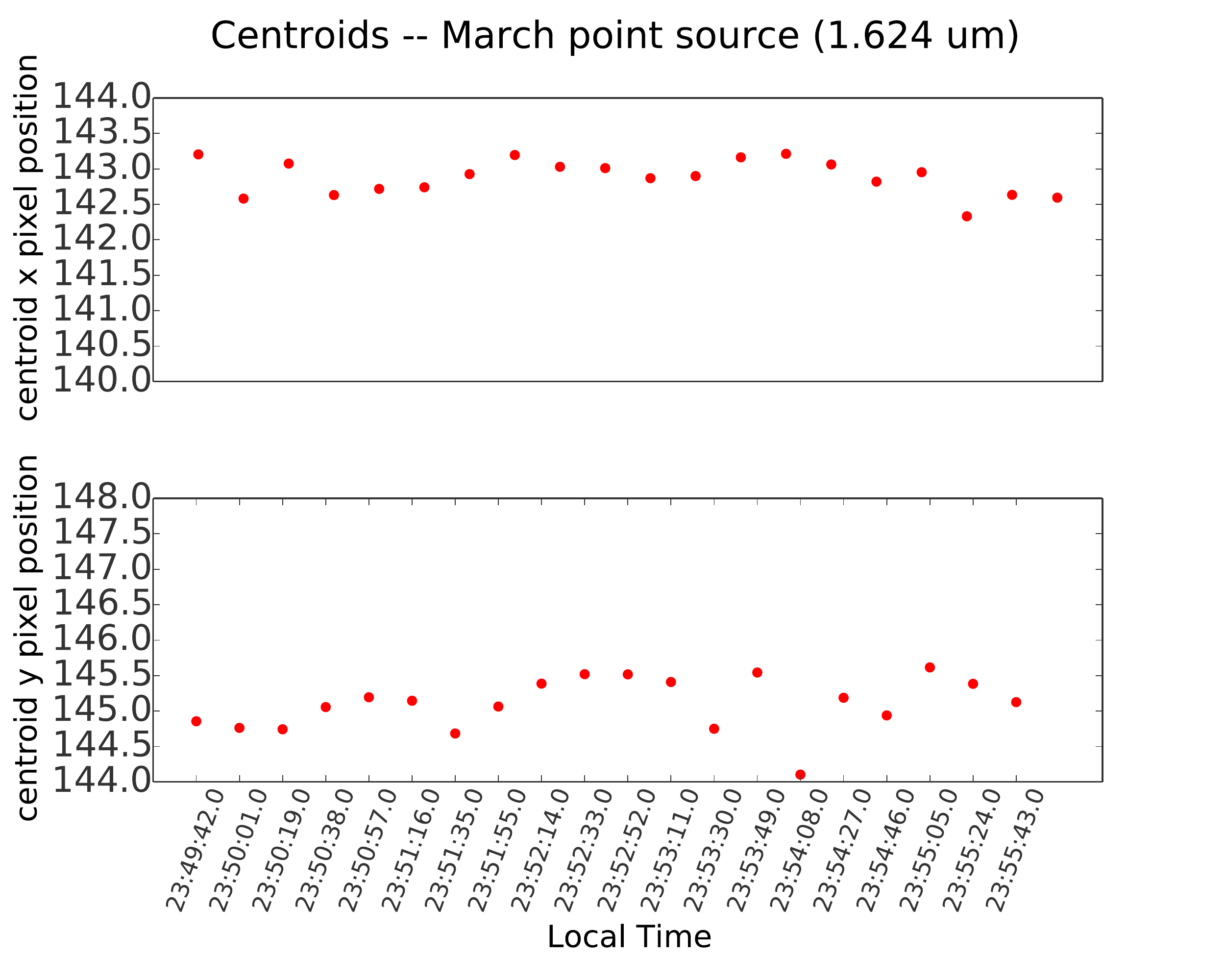}
\includegraphics[height=6.5cm]{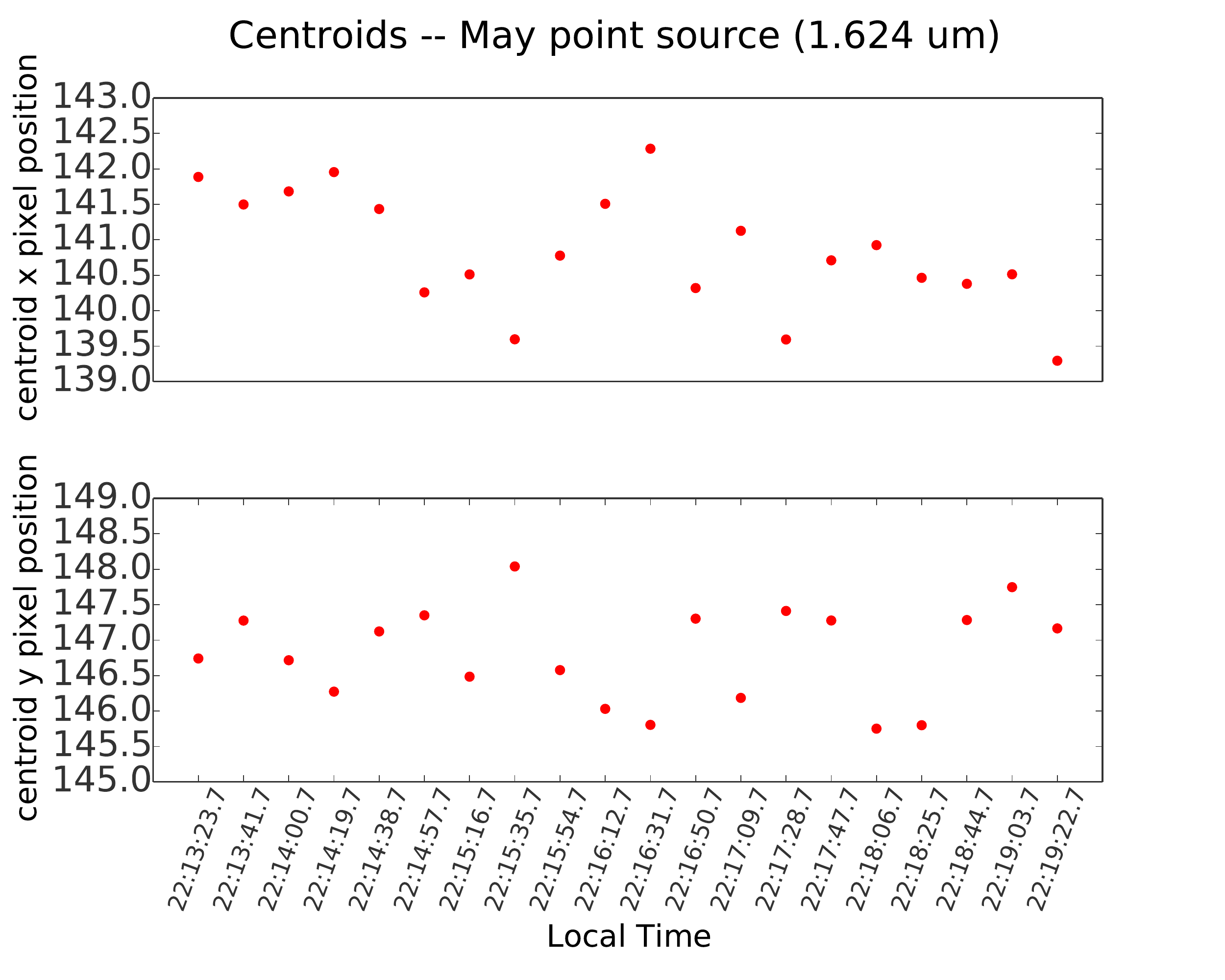}
\caption {\label{fig:centroid} \textbf{Image centroid position with time:} Both
March and May 2014 runs looked at the same point source for 1.5s exposures,
approximately 20s apart. Both show some jitter in the pointing. The pointing
stability appears worse for the May run, consistent with higher wind speeds and
poorer observing conditions. } 
\end{figure}

Instabilities in telescope pointing cause the PSF to smear during exposures,
decreasing the sensitivity and obtainable contrast of all modes. NRM is
particularly vulnerable, as analysis requires accurate measurement of fringes
which are blurred by this effect. Pointing errors causes a change in fringe
phase and a decrease in amplitude, both important quantities for NRM. This has
particular impact on the ability to measure accurate interferometric
visibilities on long exposures since different frames are affected to different
extents, making calibration difficult.
Analysis of short exposure images shows that most frames are unaffected.  This
indicates the frequency of jitter in on a several second timescale.
\begin{figure} [!htbp]
\centering \includegraphics[height=5cm]{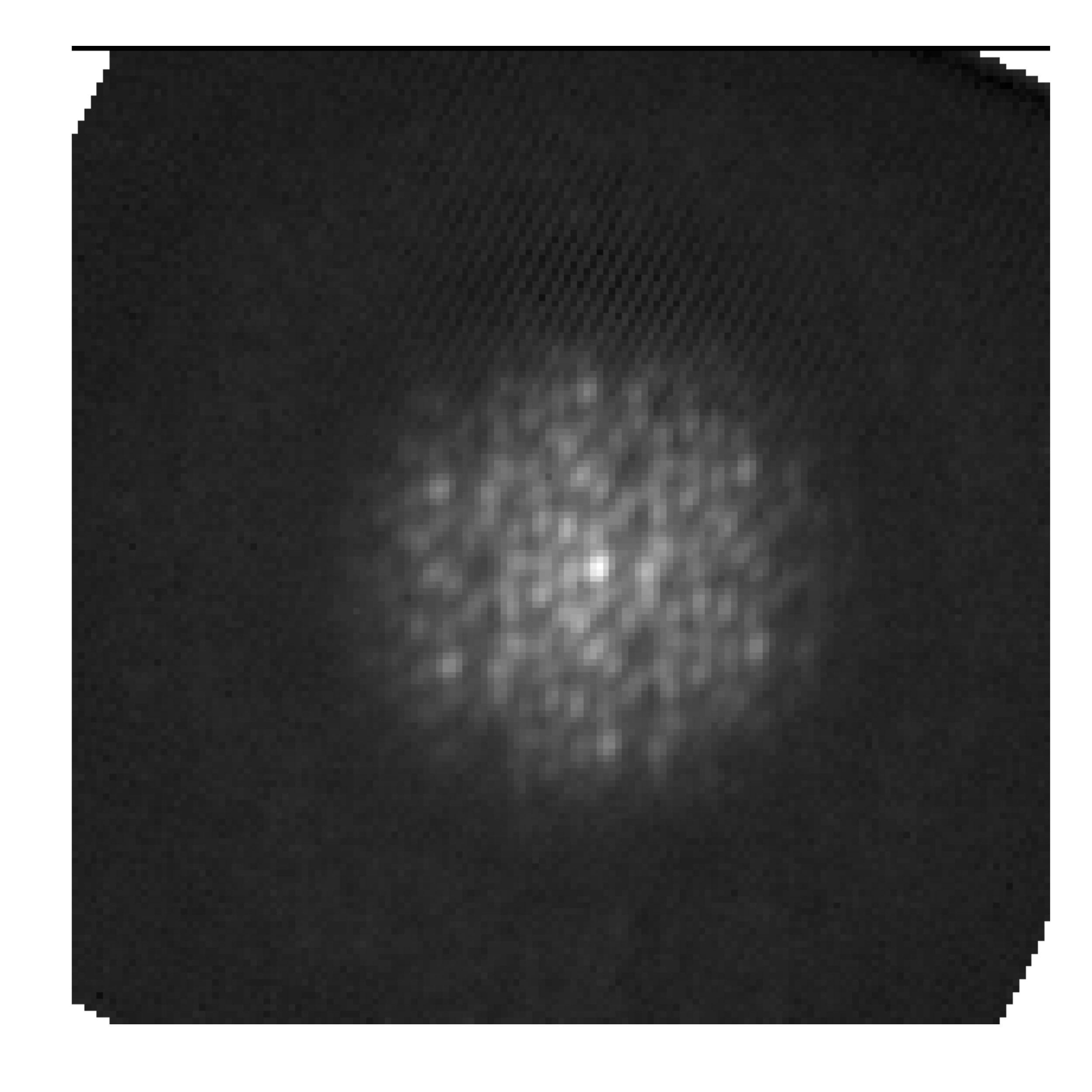}
\includegraphics[height=5cm]{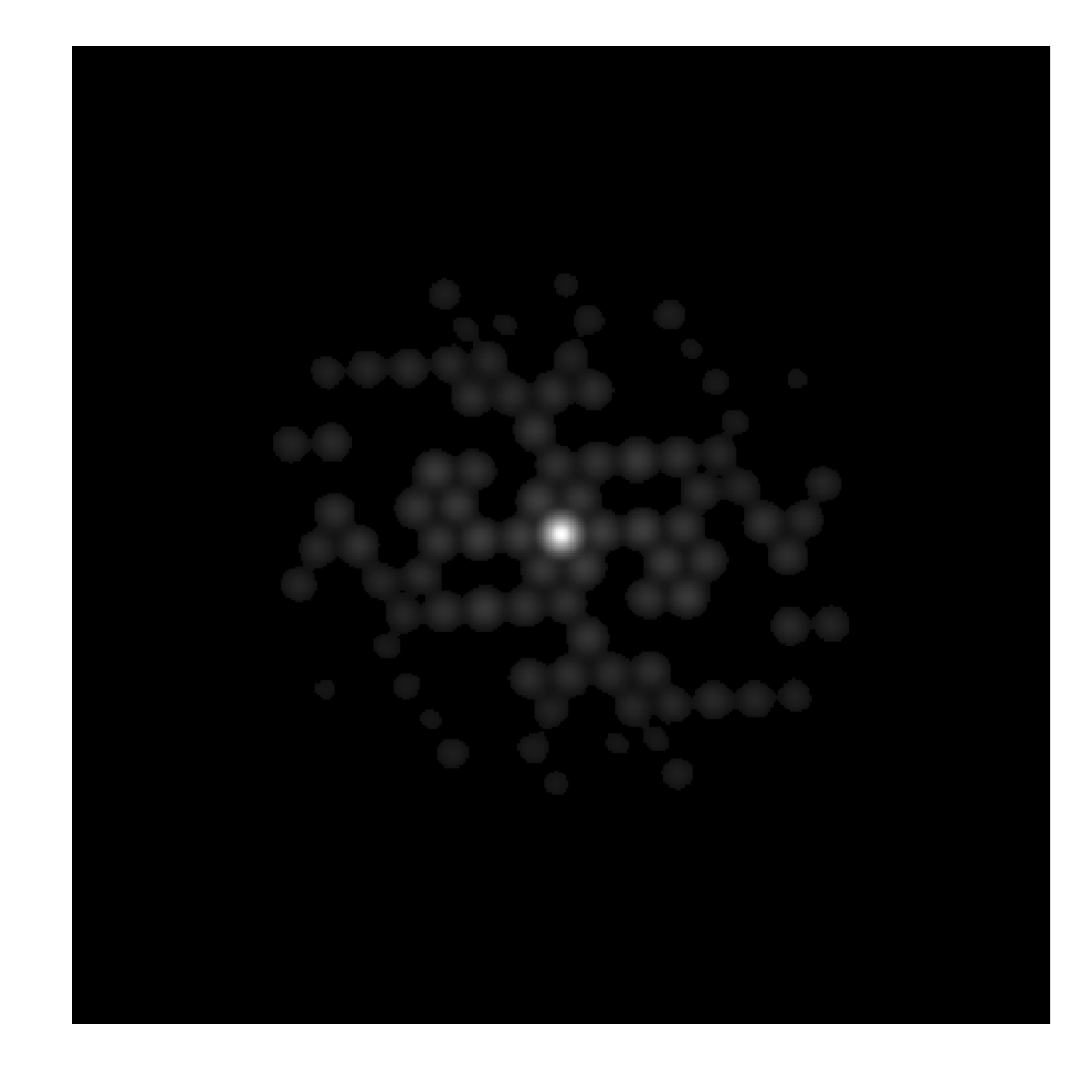}
\includegraphics[height=5cm]{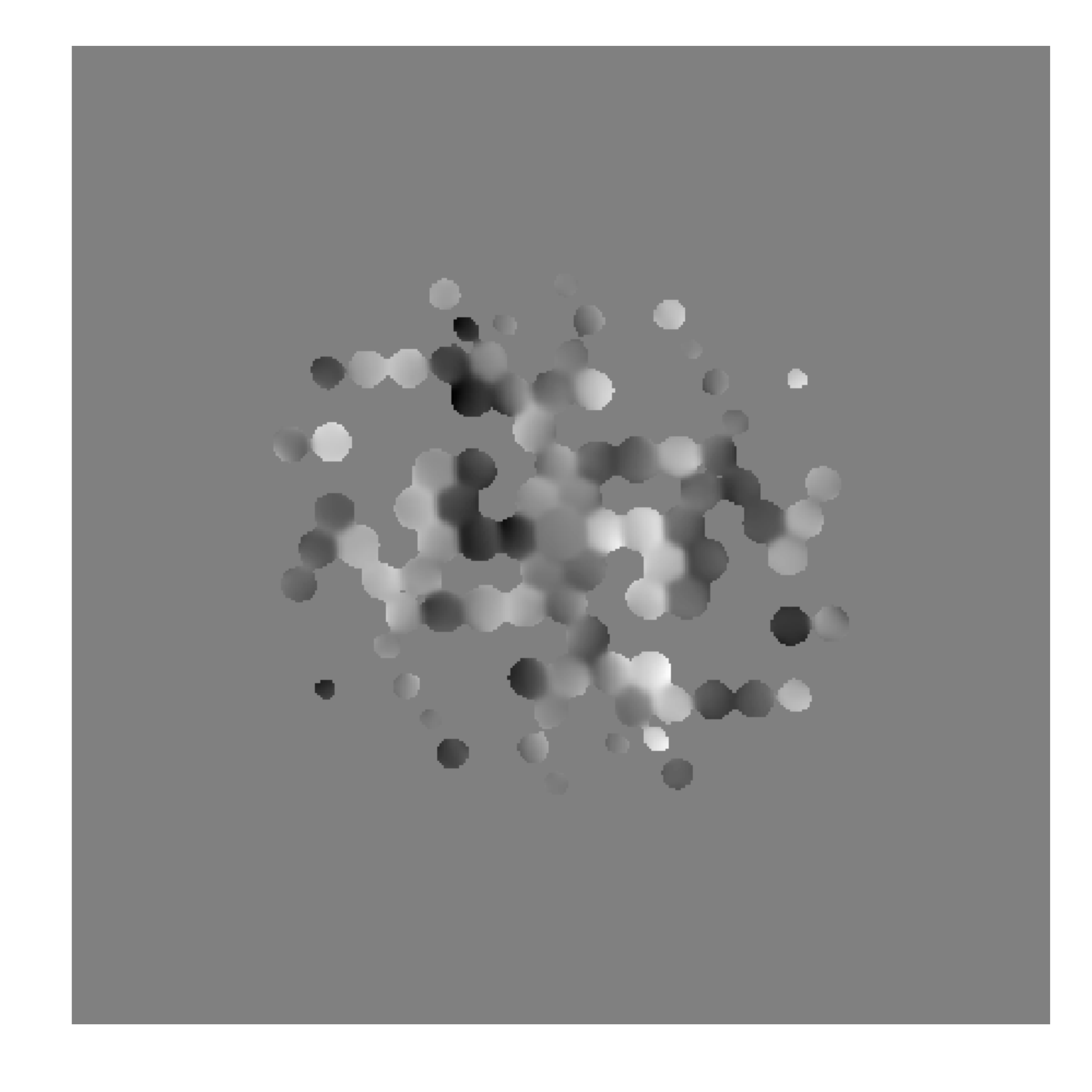} 
\caption {\label{fig:smear} \textbf{Image smearing from jitter:} \textbf{Left:}
the image shows smearing due to some source of motion. \textbf{Center:} The
power spectrum shows the fringe amplitudes of the drop significantly in
affected spatial frequencies (for example, compared to a more evenly filled
amplitude signal in Figure \ref{fig:labphases}). \textbf{Right:} Low fringe
amplitudes in turn produce lower fringe phases signal at those spatial
frequencies.} 
\end{figure}

\subsection{Large dynamic pistons}
NRM images are sensitive to systematics not seen with the coronagraph.  In the
December 2013 point source images we saw fringe jumps in the data: large and
sudden changes in the static wavefront error, causing a shift in fringe phase
and amplitude between consecutive frames of the same source. In March we saw
the same behavior return in short exposures. In the amplitudes these shifts
often appeared with clipping over particular holes. The large change in fringe
phases occurs in both short and long baselines. 

This instability was particularly pernicious given its tendency to temporarily
change the fringe visibility, making calibration difficult. There is also some
evidence to suggest that the systematic closure phases were affected, also
leading to poor calibration.

\begin{figure} \centering
\includegraphics[height=7cm]{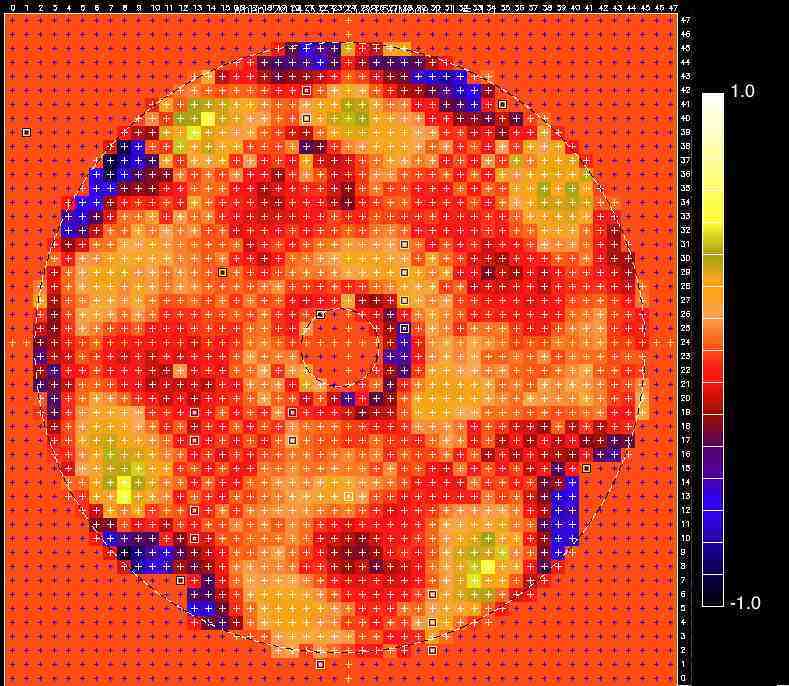}
\caption{\label{fig:tweeter} \textbf{March 2014 Telemetry:} Phase (microns)
builds on the tweeter DM trying to correct rotation seen between the WFS
lenslet and CCD pixels.} \end{figure}

One convincing explanation of big fringe jumps is a buildup of phase on the
tweeter DM.  
A rotation of the whole reference centroid pattern (e.g. due to rotation of the
lenslets relative to the CCD) should be reconstructed into a flat wavefront, as
no phase aberration upstream of the lenslets can produce that motion. However,
the GPI reconstructor projects rotation into a wavefront aberration
(concentrated at the pupil edges), due to both the Fourier Transform
Reconstruction edge extension and the missing subapertures that cover MEMS bad
actuators. Change in the wavefront sensor (WFS) gain changes the measured
rotation of the centroids relative to the (slightly rotated) reference centroid
set, producing wavefront anomalies at the pupil edge that can vary rapidly.
This was mitigated in the May 2014 run by explicitly removing the rotation
component of the centroids before reconstruction \cite{poyneerthis}.
Figure \ref{fig:tweeter} shows an example of this buildup of phase, which would
contribute phase to both short and long baselines.  These fringe jumps were not
visible in May data, though the closure phases were in general less stable with
poorer weather conditions. While more work on analyzing both the data and
telemetry is required to confirm this connection, this demonstrates NRM as a
tool for identifying systematics that are hard to diagnose with other modes.

\begin{figure} \centering
\includegraphics[width=5cm,height=5cm]{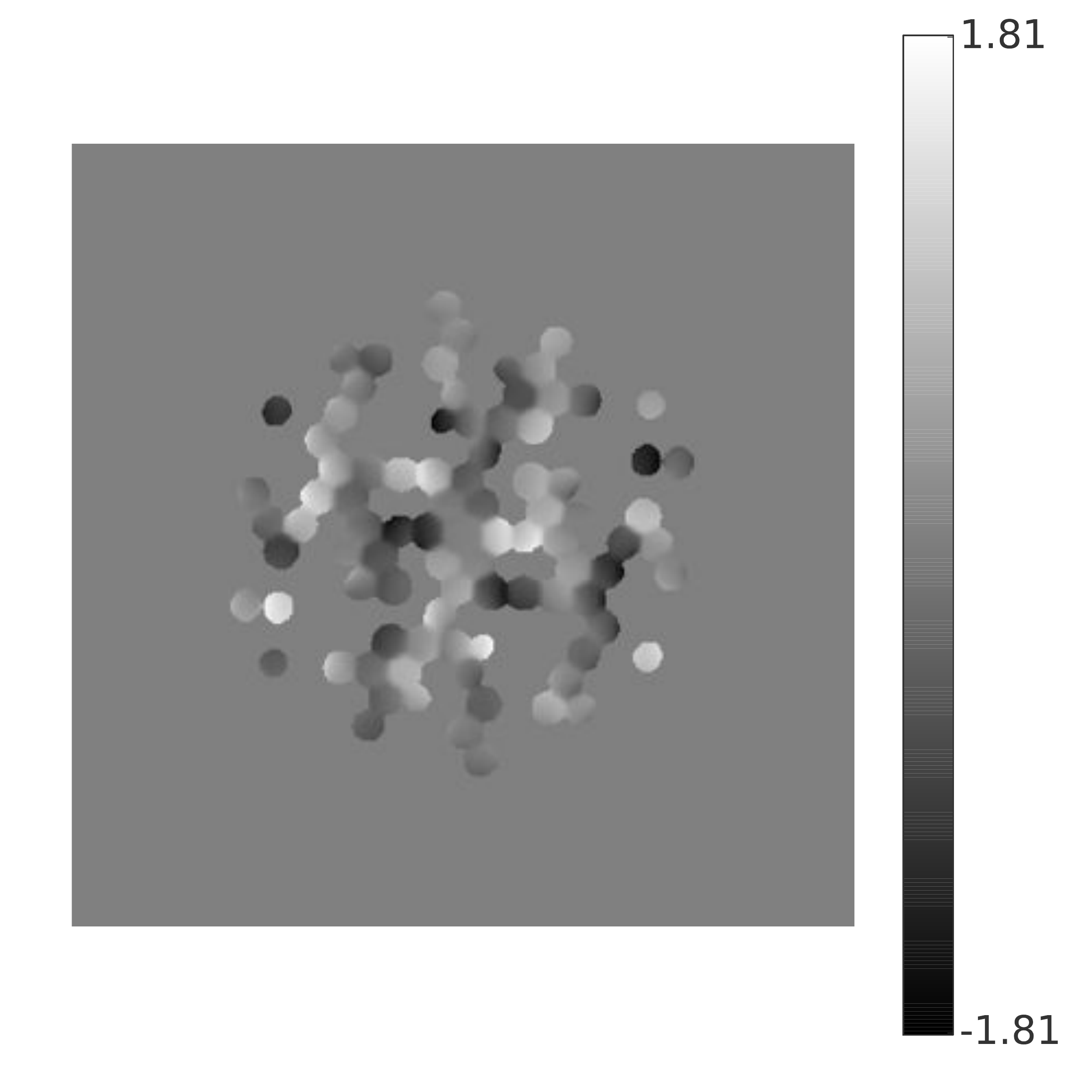}
\includegraphics[height=5cm]{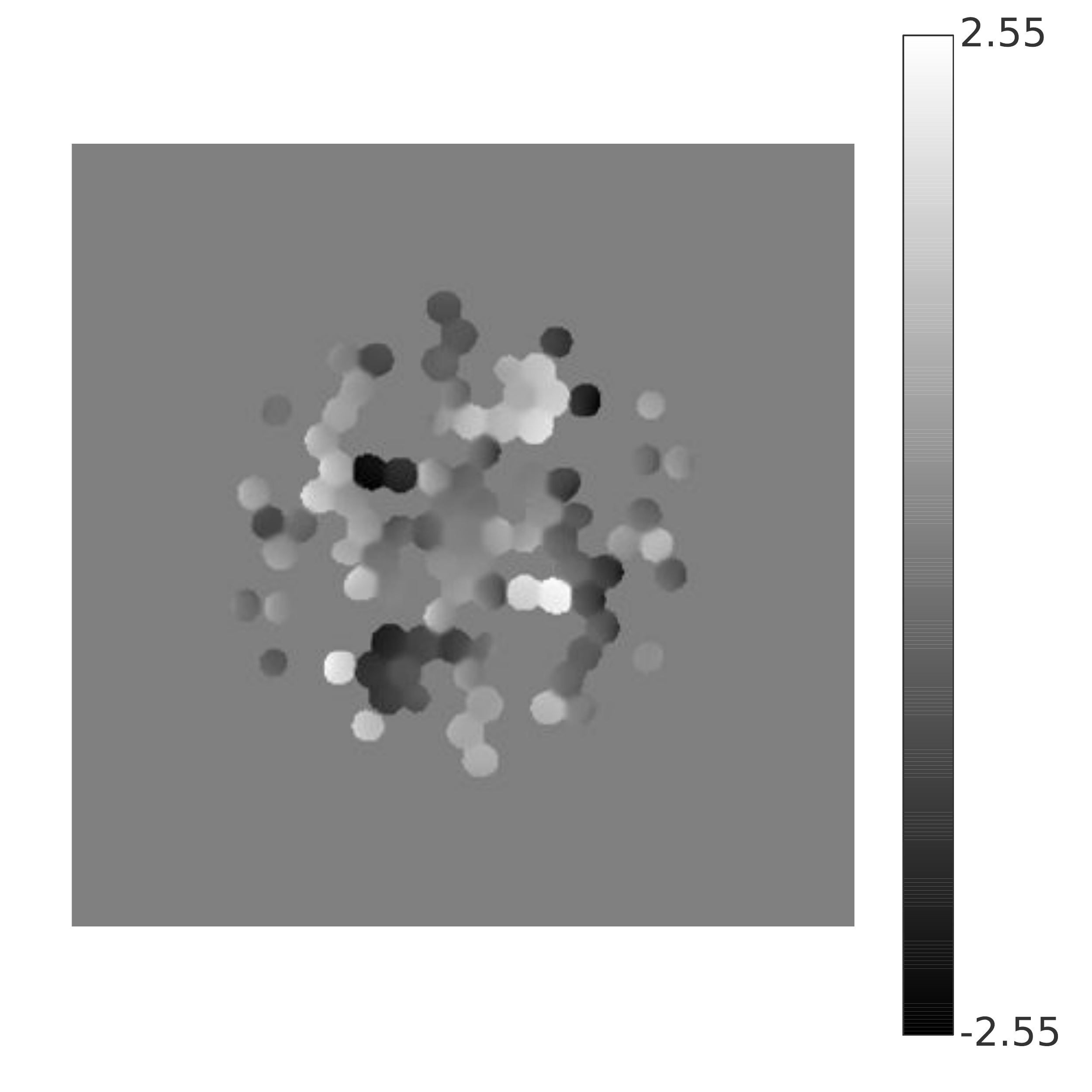} \\
\includegraphics[height=5cm]{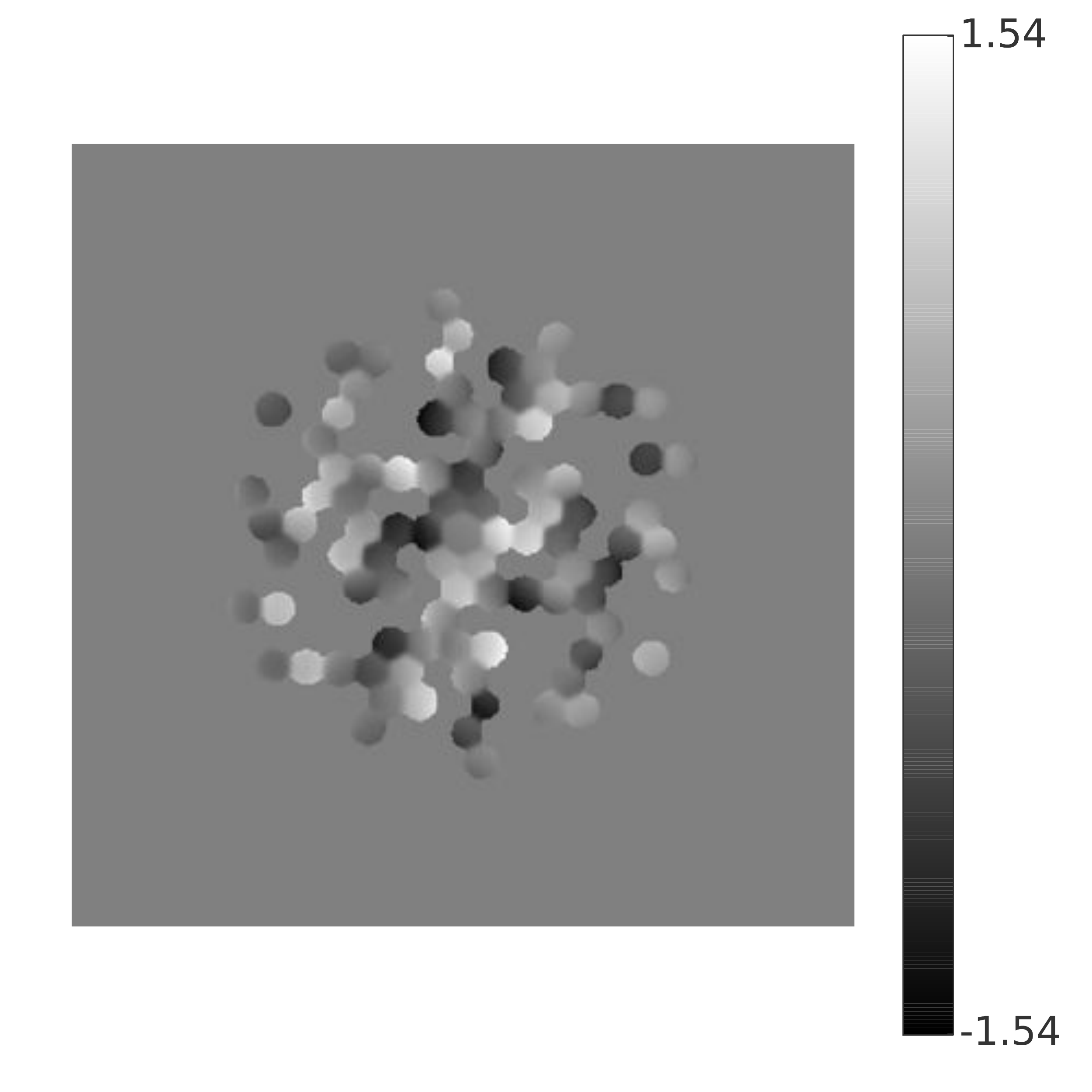}
\includegraphics[height=5cm]{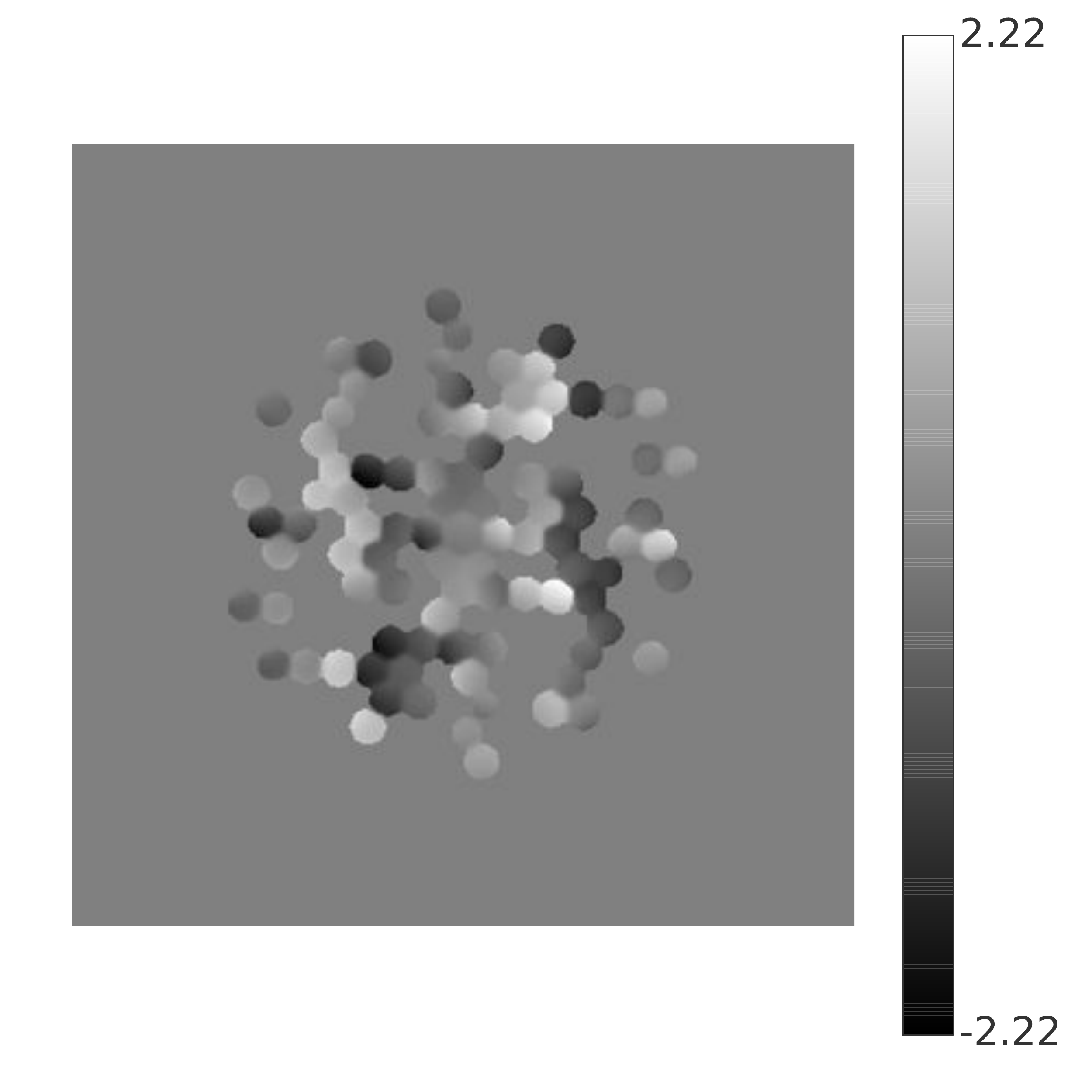}
\caption{\label{fig:dynamicphases} \textbf{Top:} Consecutive exposures of point
source HR 2716 in December 2013. \textbf{Bottom:} Consecutive exposures of HD
63852 in May 2014. Both show a similar shift in fringe phases during the
observing sequence.} \end{figure}

\section{Performance } \label{sec:performance}
\subsection{Binary detection}
As a demonstration, in December we observed known binary HR 2690 in $H$ band.
We recovered the companion at separation of $88.4 \pm 0.5$ mas and contrast
ratio of $5.94 \pm 0.09$. We recovered the companion in 24 minutes of data.
Section \ref{sec:cpstability} discusses a current estimate of detection limits
for short exposures. \\
\subsection{Closure phase stability} \label{sec:cpstability}
\begin{figure}
\centering
\includegraphics[height=5.5cm]{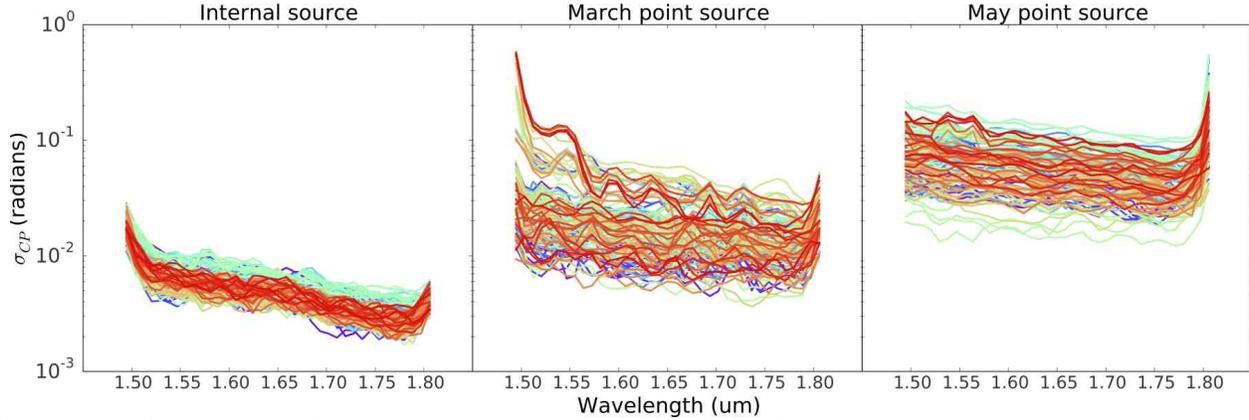}
\caption{\label{fig:cperr} \textbf{Perfomance:} Closure phase standard
deviation from internal source exposures on the telescope.  Phases are very
stable and predict raw contrast as low as  several $\times \ 10^3$ for similar
conditions. On sky in March closure phase error was roughly an order of
magnitude higher, and in May even worse. March and December had clear
conditions while May had thin and thick cirrus. We've reject 3 frames of
visible smearing in the May dataset for this calculation.}
\end{figure}

Despite a few outstanding instabilities, GPI's NRM is capable of high dynamic
range measurements, competitive with masking modes on older instruments. In
long exposures where pointing jitter smears the data, variations can be
calibrated if enough data are taken.
In the most stable conditions on the telescope from the internal source
(without atmospheric turbulence) closure phase is stable, a median standard
deviation of 0.0046 radians in $H$ band (uncalibrated).  In Figure
\ref{fig:cperr} we plot the closure phase error with wavelength for the
internal source compared with measurements on sky in March and May 2014.

In March 2014, short exposures saw relatively stable closure phases that had a
median standard deviation $0.016$ radians, uncalibrated. These closure phases
roughly follow the wavelength trend seen in July 2013 lab data
\cite{2013SPIE.8864E..1VG} and in exposures with the internal source. 
Observing conditions were worse in May and observations saw increased pointing
jitter (Figure \ref{fig:centroid}). The median standard deviation in closure
phase was 0.055 radians. This could be an explanation for larger error in the
closure phases, decreased contrast sensitivity.  In Figure \ref{fig:cperr} the
three most visibly smeared frames from are not included in analysis.

Analyzing the data through the Sydney pipeline, March 2014 short exposure
closure phases had a median standard deviation of $1.57$ degrees for each group
of 10 frames after calibration, allowing detections of point sources at up to
6.5 magnitudes. In the 17 exposures in May 2014, calibrating the first 9 frames
with the remaining frames the root-mean-square (rms) of the calibrated closure
phases was 1.62 degrees. This would provide robust detections at up to 6
magnitudes.  These results suggest the current performance of GPI NRM is
similar to that obtained routinely with older instruments such as NIRC2 at Keck
and NACO at VLT. However, the extra wavelength dimension provides more
independent data per frame than is obtained with these instruments, resulting
in higher contrast detection limits with the same number of frames and similar
closure phase scatter.

More consistent and calibratable measurements can be made with stable pointing.
When seeing and conditions are poor, it will be difficult to obtain good
contrast with NRM in the current state of the instrument. Fixing image jitter
will help bring contrast sensitivity closer to levels seen in the lab. 
We are using two independent pipelines and tracking down discrepencies in order 
to further understand systematics and how much we can calibrate them.

\section{Discussion}
As a diagnostic tool NRM can be leveraged for measuring the wavefront.
Wavefront phase reconstructions can be made from fringe phases in the data.
Comparison of analytic models with and without observed phases also provides a
good estimate for strehl ratio calculation. This would assess the quality of
the wavefront when the instrument operates without CAL corrections. In the
future, NRM may also be used to test atmospheric dispersion in IFS mode, since
the sub-pixel centering of the PSF is precisely measureable. With its finer
resolution, NRM can also be used for plate scale calibration from
known-separation binaries.

An upgrade to the real-time AO control software correction that accounts for
the rotation between CCD pixels and WFS lenslets  \cite{poyneerthis} appears
to fix the anomalous fringe jumps seen in earlier data.
While there was phase variation in May, when the new AO control software was
installed and applied, this distinct phase jump was not observed. Pointing
jitter appeared to be worst in May, when cloud cover, wind, and seeing were
worst. This was likely the biggest source of instability.  It will be difficult
to understand phase systematics without fixing the pointing jitter. A good test
in the future may be to take NRM exposures with reduced vibrations, when the
IFS cryo-coolers are turned off, and see if there are any improvements or
changes in the fringe behavior. 

Even with remaining instabilities, NRM on GPI is competitive with older NRM
instruments, and can additionally provide both spectral and polarimetric
information. While, this analysis focued on IFS data, NRM in polarization
will be a very interesting imaging mode. Solving image jitter will make
amplitude measurements more stable enabling better NRM imaging of disks. NRM in
polarization may be able to observe polarized disks closer in than other
imaging modes. Looking at large disk gaps close in will open a new search space
for GPI at the Y,J,H,K wavebands. 

While NRM is demonstrating moderate contrast capability, there remain
instrument systematics inhibiting better NRM performance. A trend of lower
amplitudes at particular longer baselines may just be a property of the
instrument, but can be manageable with a reasonable level of stability. By far
the most serious factor limiting NRM performance is image jitter on sky. This
reduces the contrast attainable for fainter targets that require longer
exposures.  While some of the effect will calibrate out, the data will lack
some visibility amplitude information at longer baselines. 

NRM is a very promising mode on GPI, for young star companion and disk science
and also instrument calibrations and diagnostics.  There are more calibration
opportunities to explore with NRM, especially those that can be done with
brighter targets. Improved wavefront control will make possible improved contrast 
detection limits as well as precision measurement on sky.


\acknowledgments The GPI project has been supported by Gemini Observatory,
which is operated by AURA, Inc., under a cooperative agreement with the NSF on
behalf of the Gemini partnership: the NSF (USA), the National Research Council
(Canada), CONICYT (Chile), the Australian Research Council (Australia), MCTI
(Brazil) and MINCYT (Argentina). This work has also been supported by NASA
grant APRA08-0117, NSF grant AST-0804417, the STScI Director’s Discretionary
Research Fund, and the National Science Foundation Graduate Research Fellowship
Program under Grant No. DGE-1232825.

\bibliography{gpinrm2014}   
\bibliographystyle{spiebib.bst}   

\end{document}